\documentclass[journal]{IEEEtran}
\pdfoutput=1

\usepackage{amsmath,amsfonts}
\usepackage{amssymb}
\usepackage{multirow}
\usepackage{array}
\usepackage{bm}
\usepackage{color}
\usepackage{booktabs}
\usepackage{subfigure}
\usepackage{textcomp}
\usepackage{stfloats}
\usepackage{url}
\usepackage{verbatim}
\usepackage{graphicx}
\usepackage{cite}
\usepackage[table,xcdraw]{xcolor}
\usepackage{algorithm}
\usepackage{algpseudocode}
\usepackage{hyperref}

\usepackage{pifont}

\newcommand{\cmark}{\textnormal{\ding{51}}}
\newcommand{\xmark}{\textnormal{\ding{55}}}

\input{mysymbol.sty}

\begin{document}

\title{Feature Selection via Graph Topology Inference for Soundscape Emotion Recognition}

\author{Samuel Rey,
    Luca Martino,
    Roberto San Mill{\'a}n-Castillo,
    and~Eduardo Morgado
\thanks{S. Rey, R. San Mill{\'a}n-Castillo, and E. Morgado are with the Department of Signal Theory and Communications, Universidad Rey Juan Carlos, Madrid, Spain. Emails: \{samuel.rey.escudero, roberto.sanmillan, eduardo.morgado\}@urjc.es.
L. Martino is with the Department of Economics and Business, Universit{\'a} degli Studi di Catania, Catania, Italy. Email:  luca.martino@unict.it}
\thanks{This work was partially supported by Agencia Estatal de Investigación-AEI, Grant PID2022-136887NB-I00 (POLIGRAPH) funded by MCIN/AEI/10.13039/501100011033, by the PIAno di inCEntivi per la RIcerca di Ateneo 2024/2026 (UPB 28722052159) and PIACERI Starting Grant BA-GRAPH (UPB 28722052144) of the University of Catania, and by the Community of Madrid within the ELLIS Unit Madrid framework and the grants URJC/CAM F1180 (CP2301) and TEC-2024/COM-89.}}

\maketitle

\begin{abstract}
Research on soundscapes has shifted the focus of environmental acoustics from noise levels to sound perception, incorporating contextual factors. 
Soundscape emotion recognition (SER) models perception using a set of features, with arousal and valence commonly regarded as sufficient descriptors of affective states. 
In this study, we blended \emph{graph learning} techniques with a novel \emph{information criterion} to develop a feature selection framework for SER. 
Specifically, we estimated a sparse graph representation of feature relations using linear structural equation models (SEM) tailored to the widely used Emo-Soundscapes dataset. 
The resulting graph captures the relationship between the input features and the two emotional outputs. 
To determine the appropriate level of sparsity, we propose a novel \emph{generalized elbow detector} that provides both a point estimate and an uncertainty interval. 
We conducted an extensive evaluation of our methods, including visualizing the inferred relations. 
While several of our findings align with previous studies, the graph representation also reveals a strong connection between arousal and valence, challenging the common SER assumptions.
\end{abstract}


\section{Introduction}
\IEEEPARstart{I}{n} recent years, soundscape research has shifted its focus in environmental acoustics from noise levels to how communities and individuals perceive sounds, incorporating contextual factors. 
The term soundscape, coined by Southworth and later popularized by Schafer~\cite{southworth1967sonic,schafer1993soundscape}, encompasses a broad range of applications, including urban sound planning~\cite{van2020multi}, acoustic monitoring~\cite{segura20215g}, sound design~\cite{gorne2019emotional}, and sonification~\cite{abri2021comparative}.
Soundscape emotion recognition (SER) is an active research area that addresses challenges such as the definition of perceptual descriptors~\cite{mitchell2024soundscape}, probabilistic modelling approaches~\cite{mitchell2022analyse}, and the influence of psychological well-being and demographic factors~\cite{erfanian2021psychological}.
Compared to the emotions elicited by music or speech, those associated with soundscapes are subtler and are influenced by the complex interplay of physical, social, cultural, and psychological factors.
{Generally, music and speech are intentionally structured to arouse a range of emotions, which are reflected in specific signal attributes~\cite{fiebig2020assessments}. In contrast, soundscapes are rarely designed to induce emotions, and the features that drive emotional responses cannot be directly inferred from studies on speech and music emotion~\cite{ma2015human}.}
Among the various models used to represent this complexity, Russell’s circumplex model~\cite{russell1980circumplex,axelsson2010principal,davies2014soundscape,fan2017emo} has been widely adopted, focusing on two affective dimensions, arousal (i.e., eventfulness) and valence (i.e., pleasantness), which are considered sufficient descriptors of perceived affect in soundscape studies.
Modelling SER offers an efficient alternative to perceptual surveys and supports a deeper understanding of soundscape perception.
While nonlinear models often outperform linear models, the latter remain attractive because of their simplicity~\cite{lionello2020systematic}. 
However, SER predictive frameworks vary considerably in terms of the choice of datasets and descriptors.

In the last decade, the scientific community has made an effort to offer valuable open-source databases, including general-purpose collections such as Emo-Soundscapes~\cite{fan2017emo}, urban soundscape datasets such as {ARAUS~\cite{ooi2023araus},} ATHUS~\cite{giannakopoulos2019athens} and ISD~\cite{mitchell2022analyse}, and indoor collections such as HSDD~\cite{versumer2023extensive}.
In this study, we focus on the Emo-Soundscapes (EMO) database, which is the most widely explored dataset in SER research~\cite{abri2020predicting,abri2021comparative,fan2017emo,fan2018soundscape,ntalampiras2020emotional,OurPaperSound,krishan2022classifying,VariableTecnia2022,san2024variable,serradilla2024emotional,manivannan2024emotioncaps}.
{Moreover, we extend our analysis to the ARAUS dataset, which employs alternative experimental setups and emotion rating protocols to EMO. This provides additional dataset diversity to assess the robustness of our feature selection strategy for SER.}
Identifying the features that most strongly influence perceptual descriptors remains an open research problem, which motivates our focus on feature selection for the EMO database. 
Existing approaches to feature selection fall into three main categories: (a) wrapper methods, which sequentially evaluate subsets of variables by optimizing a cost function; (b) filter-based methods, which rank variables based on statistical criteria independently of the model; and (c) embedded methods, which integrate the selection into model training.
{A considerable number of studies have explored diverse variable selection strategies grounded in EMO for regression tasks. In some cases, the ranking stage was skipped~\cite{fan2018soundscape,ntalampiras2020emotional,serradilla2024emotional}, while other studies focused on iterative feature elimination approaches~\cite{abri2021comparative,OurPaperSound}. Classical techniques such as principal component analysis (PCA)~\cite{abri2020predicting} and Gini Importance~\cite{san2024variable} have also been employed to rank variables. 
Regarding the selection of the final subset of variables, several studies rely on heuristic procedures or model-based criteria to determine the retained features~\cite{ntalampiras2020emotional,fan2017emo}. For instance, linear regression models have been used to identify the most influential predictors~\cite{abri2020predicting}, whereas hyperparameter grid searches for random forests have been explored to determine the best-performing feature subsets~\cite{abri2021comparative}. Decision tree-based models have also been considered to obtain parsimonious predictors~\cite{san2024variable}. In addition, recent works have explored alternative representations based on autoencoders for modeling soundscape emotions~\cite{serradilla2024emotional}.}
{The literature also reports variable selection analyses conducted on alternative datasets. In~\cite{orga2021multilevel}, the authors implemented a backward stepwise selection approach using an urban traffic noise dataset. A related strategy was adopted in~\cite{mitchell2021investigating}, where the Akaike Information Criterion (AIC) was used to evaluate the ISD dataset, which comprises 11 original variables.}

In this context, graph-based representations offer an alternative framework for capturing the pairwise relationships among variables. 
By encoding interactions as edges between nodes, graphs can represent a variety of underlying structures in the data, ranging from statistical dependencies to perceptual associations~\cite{kolaczyk2009book,shuman2013emerging}. 
Relating the properties of observed signals with the topology of the graph lies at the core of graph signal processing (GSP), a field devoted to learning from data defined over irregular domains modeled by graphs~\cite{ortega2018graph,dong2020graph,leus2023graph}. 
A central problem in GSP is \emph{graph learning}, which seeks to infer a graph structure from observed signals that captures the underlying relations among variables~\cite{lake2010discovering,egilmez2017graph,mateos2019connecting,navarro2022joint}. 
The key to this task is leveraging signal models that connect the signal properties with the graph topology.
Notable approaches include partial correlations and Gaussian graphical models~\cite{friedman2008sparse,egilmez2017graph,rey2023enhanced}, smoothness-based models~\cite{kalofolias2016learn,dong2016learning,buciulea2021learning}, and graph-stationary models~\cite{segarra2017network,shafipour2020online,navarro2024joint,buciulea2025polynomial}. 
{Of particular interest are structural equation models (SEM), which assume that the data follow a linear autoregressive process governed by the graph topology, where each node is expressed as a linear combination of its neighbors~\cite{cai2013sparse,baingana2014proximal,giannakis2018topology}. 
SEM provides an explicit parametric interpretation of edge weights as direct influence coefficients and naturally accommodates sparsity-promoting regularization. 
Moreover, SEM-based graph learning has attracted increasing attention beyond signal processing due to its ability to model potentially causal relationships~\cite{zheng2018dags,yang2022causal,wang2023causal,montagna2023scalable,rey2025non}. 
These properties render SEM particularly appealing for modeling the underlying structure of soundscape emotion data.}

{The use of graph-based representations is gaining attention in soundscape analysis and emotion recognition tasks. 
For instance,~\cite{gao2024adaptive} constructs graphs over short time frames within an audio segment based on the similarity between frames for speech emotion classification. In~\cite{nie2020c}, nodes correspond to videos and edges encode inter-video relations, and the resulting graph is used for emotion recognition.
Similarly,~\cite{li2023graphcfc} models conversational flows by representing utterances as nodes and their interactions as edges, while~\cite{hou2023audio} builds graphs over detected audio events to support an explainable acoustic scene classification.
In addition,~\cite{yin2023robust,yin2025leveraging} employed correlation graphs for feature selection in streaming settings. 
In a different context, modeling relationships among samples using graphs and higher-order structures has also gained traction in feature selection~\cite{sheikhpour2025semi,sheikhpour2025sparse}.
Table~\ref{TableResultsLiterature2} summarizes related works, highlighting differences in the ranking method, the choice of the number of variables, the use and type of graph, and whether the graph topology is formally inferred.}


{In contrast to prior approaches, which construct graphs from heuristic similarity measures rather than formally inferring the underlying topology of the data, our work aims to infer a graph over acoustic features through a graph learning framework. Rather than modeling the relationships between segments, videos, or events, we focused on uncovering structural dependencies among features to support principled feature selection in soundscape emotion recognition. 
In particular, this work integrates techniques from graph learning and information criteria to develop a novel variable selection approach for SER.}
More specifically, we assume that every variable can be expressed as a linear combination of its neighbors, following a linear SEM~\cite{kaplan2008structural,shimizu2011directlingam}.
In contrast to previous methods, the two outputs (arousal and valence) were considered jointly with the inputs. 
This modeling approach leads to a convex optimization problem for inferring the underlying graph topology, where we include a regularization term to promote sparsity. 
While most graph learning methods treat the sparsity regularization parameter $\lambda$ as a tunable hyperparameter, we propose a novel and systematic alternative for selecting its optimal value.
We introduce a generalized version of the automatic elbow detector, which can be interpreted as a modern information criterion~\cite{UAEDmorgado2023universal}. 
In addition to estimating the optimal $\lambda$, this generalization yields an interval-based solution that provides an uncertainty estimate. {The differences with respect to previous work are highlighted in Table~\ref{TableResultsLiterature2}.}

In summary, the proposed approach can be interpreted as a combination {of} wrapper and filter methods for feature selection. Specifically, we adopted a graph topology learning framework that enables the joint analysis of the dependencies and connections among all variables, including both inputs and outputs. This allows us to quantify the strength of connections with the outputs (as in a wrapper method, facilitating the construction of rankings) and identify and discard variables based on their correlations (as in a filter method).

The remainder of this work is organized as follows.
{Section~\ref{s:databases} describes EMO and ARAUS databases.}
{Section~\ref{s:graph_learning} presents the proposed graph learning framework, and Section~\ref{s:elbow_detector} details how to select the regularization parameter.}
Section~\ref{s:experiments} presents and discusses the results of several experiments. 
Finally, Section~\ref{s:conclusions} summarizes the main findings and conclusions and outlines future research directions.

\begin{table*}[t]
\begin{center}
\caption{Summary of related works in the literature.} 
\label{TableResultsLiterature2}

\vspace{-0.1cm}
{
\begin{tabular}{c l c c c c} 
\toprule
Work & Ranking Method & Choice of number of variables & Graph-based & Topology inferred & Nodes \\ 
\midrule

\cite{fan2017emo} & Variance & Heuristic Threshold & \xmark & \xmark & -- \\ 
\cite{fan2018soundscape} & No ranking & Heuristic & \xmark & \xmark & -- \\ 
\cite{abri2020predicting} & Principal Components Analysis (PCA) & Linear Regression Test (KBest) & \xmark & \xmark & -- \\ 
\cite{abri2021comparative} & Recursive Feature Elimination & Grid Search (wrapper method) & \xmark & \xmark & -- \\ 
\cite{ntalampiras2020emotional} & No ranking & Heuristic & \xmark & \xmark & -- \\ 
\cite{OurPaperSound} & Forward Feature Selection \& others & Heuristic by Gibbs analysis & \xmark & \xmark & -- \\ 
\cite{DTR_Access_24} & Gini Importance & SIC -- UAED - CV procedure & \xmark & \xmark & -- \\ 
\cite{serradilla2024emotional} & No ranking & Auto-encoders & \xmark & \xmark & -- \\ 

\midrule

\cite{yin2023robust} & Fuzzy dependency + PageRank centrality & Top-k features & \cmark & \xmark & Features \\ 
\cite{yin2025leveraging} & Manifold correlation / separability score & Top-k features & \cmark & \xmark & Samples \\ 
\cite{sheikhpour2025semi} & Projection weights with $\ell_{2,1}$ regularization & Top-k features & \cmark & \xmark & Samples \\ 
\cite{sheikhpour2025sparse} & Sparse discriminant projection & Top-k features & \cmark & \xmark & Samples \\ 

\midrule

\textbf{Here} & Structural Equation Model (SEM) & Generalized UAED & \cmark & \cmark & Features \\ 

\bottomrule
\end{tabular}
}
\end{center}
\vspace*{-.3cm}
\end{table*}

\section{Databases and scopes}\label{s:databases}

\subsection{EmoSoundscapes dataset}\label{s:emosoundscape}
This work explores the EMO dataset\footnote{\url{https://metacreation.net/emo-soundscapes/}}, a large, publicly available collection of soundscapes with emotion annotations, which, to the best of our knowledge, is the most extensively benchmarked dataset of soundscapes~\cite{fan2017emo}. 
It contains 1213 audio files released under a Creative Commons license, created by mixing sounds from the Freesound collaborative platform~\cite{fonseca2017freesound}.
For each clip, EMO provides 122 input variables that are diverse and meaningful from an acoustic signal perspective, and two output variables representing emotional labels. 
This variety of soundscapes and features provides EMO with a dimensionality that is well suited for evaluating feature selection methods.
The EMO dataset classifies soundscapes within six of Schafer's taxonomy categories~\cite{schafer1993soundscape}: human sounds (e.g., shouts and whispers), natural sounds (e.g., rain and birds), quiet and silence (e.g., silent forests and quiet parks), sounds and society (e.g., concerts and stores), mechanical sounds (e.g., factories and engines), and sounds as indicators (e.g., clocks and church bells).
These categories involve both the sound source and the listening context, providing a more general and straightforward approach than other taxonomies, such as Brown's~\cite{brown2010soundscapes}.
The dataset includes 600 clips with 100 examples per category, plus an additional 613 clips created by manually mixing and gain-adjusting sounds from two or three categories, resulting in 1213 soundscapes in total. 
Perceived emotions (arousal and valence) were obtained through a crowdsourcing procedure involving 1182 trusted annotators who completed a parallel ranking-based questionnaire consisting of pairwise comparisons between two audio clips. 
The collected ratings achieved adequate intersubject reliability.

\begin{figure}[h!]
    \centerline{
    \includegraphics[width=.7\columnwidth]{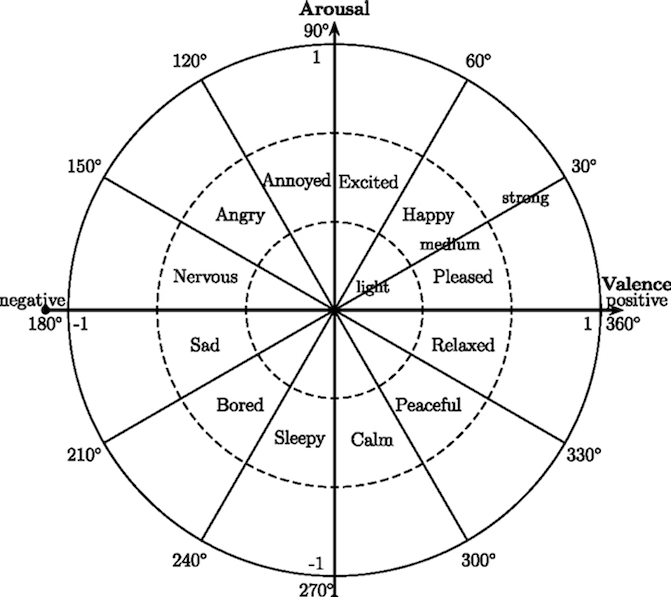}
    }
    \caption{Circle of emotion with Arousal and Valence in two cardinal positions~\cite{salmeron2012fuzzy}. 
    }
    \label{Russell}
    \vspace*{-0.3cm}
\end{figure}

The EMO files are 6s long, monophonic, and sampled at 44100Hz. 
Previous studies~\cite{xu2019soundscape,fan2016automatic} have shown that this format is adequate for assessing soundscape affective dimensions such as eventfulness (i.e., arousal) and pleasantness (i.e., valence); see Figure~\ref{Russell} for a visual representation. 
From each file, EMO extracts 122 normalized features using a 23~ms Hanning window with 50\% overlap, computed with general-purpose tools such as YAAFE~\cite{mathieu2010yaafe} and MIRToolbox~\cite{lartillot2008matlab}, {see Table~\ref{TableVariables}.}

\begin{table*}[h!]
\centering
\caption{Summary and classification of acoustic variables in the EMO and ARAUS datasets. Psychoacoustic variables refer to the perceptual (i.e., subjective) attributes of sounds. Frequency-domain variables focus on the spectral shape and harmonic structure of sounds. Time-domain variables describe the signal dynamics.} 

\label{TableVariables}

\vspace{-0.2cm}
{
\begin{tabular}{|p{2.2cm}|p{5.2cm}|p{4.0cm}|p{4.0cm}|} 
\hline
\textbf{Dataset} & \textbf{Psychoacoustic} & \textbf{Time-domain} & \textbf{Frequency-domain} \\ 
\hline

EMO &
Loudness, MFCCs, Sharpness, Fluctuation Strength (4, 24–49, 113, 114, 117–119) &
Energy, En (1–7, 22, 23, 52, 115, 116) &
Pitch, Chromagrams, Inharmonicity, Roll-off (The remainder) \\
\hline

ARAUS &
Loudness, Sharpness, Fluctuation Strength, Roughness, Tonality (1-43,76-104) &
$L_Aeq$, $L_Ceq$, percentiles (44-75, 144, 145)  &
Spectral powers (105-141), $L_Aeq$, $L_Ceq$ (44-75, 144, 145) \\
\hline

\end{tabular}
}
\vspace*{-.2cm}
\end{table*}

{\subsection{ARAUS dataset}\label{s:araus}
This study also examined the ARAUS dataset, a large-scale, publicly available collection of affective responses to augmented urban soundscapes, designed explicitly as a benchmark for perceptual soundscape research and for modeling the impact of added sound sources on human appraisal~\cite{ooi2023araus}. The ARAUS comprises 25,440 audio-visual stimuli generated by combining real-world recordings of urban environments with digitally added maskers (birds, water, wind, traffic, construction, or silence) mixed at soundscape-to-masker ratios of $-6$, $-3$, $0$, $+3$, or $+6$~dB. The resulting corpus spans a diverse range of acoustic scenes and masking conditions.
The dataset was built using 234 base urban soundscapes recorded primarily in the Urban Soundscapes of the World (USotW) database~\cite{de2017urban}, supplemented with 280 single-channel masker recordings. All audio is presented in 30-second binaural form and paired with a synchronized video crop at 0$^\circ$ azimuth and 0$^\circ$ elevation.

Each stimulus is annotated through the ISO/TS 12913-2:2018 Method A questionnaire, collecting ratings for pleasantness, eventfulness, vibrancy, calmness, monotony, chaos, annoyance, and audiovisual appropriateness~\cite{ISO_Soundscape2014part}. The final dataset incorporates responses from 605 validated participants recruited under controlled laboratory conditions. Each participant was screened using a hearing test and a battery of seven ISO-derived internal consistency checks. In addition to their perceptual assessments, the participants provided detailed demographic information.
The ARAUS dataset also provides a set of 143 features, which are summarized in Table~\ref{TableVariables}.
}

\section{Soundscape Representation via Graph Learning}\label{s:graph_learning}
We aim to learn a graph representation that models the relations among the acoustic features in the EMO dataset.  
The resulting graph captures the underlying structure of the dependencies between features, offering new insights into their interactions.  
We further leverage this learned structure to identify which features are most influential in predicting  perceived emotional responses, namely arousal and valence.  
To this end, we first present the fundamentals of graph learning and then tailor this framework to the specific characteristics of the EMO dataset.  
Finally, while our analysis and methods focus on the EMO dataset, the proposed methods are general and can be applied to other soundscape datasets as well.

\subsection{Fundamentals of Graph Learning}

A graph is a mathematical entity that encodes pairwise interactions.
Formally, a graph $\ccalG = (\ccalV, \ccalE)$ is composed of $N$ nodes collected in the set $\ccalV$, and a set of edges $\ccalE \subseteq \ccalV \times \ccalV$.  
The connectivity of $\ccalG$ is encoded in a sparse adjacency matrix $\bbA \in \mathbb{R}^{N \times N}$, where $A_{ij} = 0$ if and only if $(i,j) \notin \ccalE$, and each nonzero entry $A_{ij}$ represents the weight of the edge between nodes $i$ and $j$~\cite{shuman2013emerging}.  
An essential component in this setting are \emph{graph signals}, which are signals defined over the set of nodes $\ccalV$ and represented as a vector $\bbx \in \mathbb{R}^N$, where $x_i$ denotes the signal value at node $i$.

{Graph learning is an inverse problem aiming to estimate the graph topology, typically represented by the adjacency matrix $\bbA$, from a set of nodal observations~\cite{mateos2019connecting,rey2025online}.}
Let the matrix $\bbX := [\bbx_1, \dots, \bbx_M] \in \mathbb{R}^{N \times M}$ be a collection of $M$ graph signals whose (statistical) properties are assumed to be intimately related to the unknown graph $\ccalG$.
{Under this assumption, the graph topology can be recovered by fitting a model that links the observed signals $\bbX$ with the unknown adjacency matrix $\bbA$.}
{Then, the adjacency matrix $\bbA$ can be estimated by solving the following optimization problem}
\begin{equation}\label{e:general_nti}
    \hbA = \argmin_{\bbA \in \ccalA} L(\bbA, \bbX) + \lambda \| \bbA \|_1,
\end{equation}
where $L(\bbA, \bbX)$ is a loss function that captures the relationship between $\bbA$ and the observed signals $\bbX$, while $\| \bbA \|_1$ denotes the $\ell_1$ norm of the vectorized matrix $\bbA$, included to promote sparsity.
The parameter $\lambda \geq 0$ controls the level of sparsity, and the constraint set $\ccalA$ ensures that $\bbA$ corresponds to a valid adjacency matrix.

{The success of graph learning critically depends on the nature of the dependency between the observed signals $\bbX$ and the unknown topology encoded in $\bbA$, which is modeled through the loss function $L(\bbA,\bbX)$.}
Different approaches leverage different types of signal-structure relationships, with prominent examples including graphical models~\cite{friedman2008sparse,ravikumar2011high}, graph smoothness~\cite{kalofolias2016learn}, and graph stationarity~\cite{segarra2017network}.  
Next, we focus on estimating the graph topology under the assumption that the observed signals follow a SEM, which constitutes the setting of interest for this work.

\subsection{Learning a Graph Representation for {Soundscape} Datasets}
According to the description of the EMO dataset in Section~\ref{s:emosoundscape}, modeling the relations between the different soundscape features as a graph entails learning the topology of a graph with $N=124$ nodes, where each node corresponds to a particular feature, including the output features, arousal, and valence. 
To this end, we have access to $M=1213$ different soundscapes. 
{Similarly, when considering the ARAUS dataset, the resulting graph contains $N=145$ nodes, with $M=20022$ observations available. In both cases, the observations are collected in the data matrix $\bbX \in \reals^{N \times M}$, whose dimensions depend on the dataset under analysis.}

In the context of soundscape emotions, there is a growing interest in predicting a subset of features, particularly those related to perceived emotions, such as arousal and valence, from the remaining features. 
Based on this observation, we postulate that the observed features follow a linear SEM~\cite{kaplan2008structural,rey2025non}, meaning that the matrix $\bbX$ is given by the following autoregressive model
\begin{equation}\label{e:sem}
    \bbX = \bbA\bbX + \bbW,
\end{equation}
where $\bbW$ denotes the exogenous input matrix, whose columns are independent realizations of a multivariate distribution with a diagonal covariance. 

Here, $\bbA$ denotes the unknown adjacency matrix encoding the relations between the different features, which is assumed to be sparse and have zeros on its diagonal. 
{Equation~\eqref{e:sem} states that the value of the $i$-th feature in a given observation depends linearly on the features of its neighbors. Consequently, SEMs are particularly well suited for feature selection, as the adjacency matrix directly quantifies the predictive influence between variables.}
{Unlike conditional-independence or smoothness-based graph learning methods~\cite{friedman2008sparse,kalofolias2016learn}, SEM provides a predictive interpretation of edges,well suited to SER setting.}
Moreover, the estimation error is determined by the (unknown) covariance of exogenous input $\bbW$.

Building on the SEM in \eqref{e:sem}, a common approach to estimate the adjacency matrix $\bbA$ from the signals $\bbX$ is to solve the following convex regularized least-squares problem
\begin{alignat}{2}\label{e:sem_graph_learning}
     \!\!&\! \hbA_\lambda = \argmin_{\bbA} \
    && \| \bbX - \bbA\bbX \|_F^2 + \lambda\| \bbA \|_1  \nonumber \\ 
    \!\!&\! \mathrm{s.t.} && 
    \bbA \in \ccalA := \{ A_{ii} = 0, \;\; \bbA = \bbA^\top \}.
\end{alignat}
The first term in the objective promotes fidelity to the data, while the second term, weighted by $\lambda$, encourages sparsity in the adjacency matrix via the $\ell_1$ norm, similar to Lasso regression settings. 
{It should be noted that the adjacency matrix ${\bf A}$ is required to satisfy the following constraints:}
\begin{enumerate}
    \item  zeros on its diagonal, i.e., $A_{ii} = 0$.
    \item and it is symmetric\footnote{Note that, setting $\lambda=0$, and using a simple least squares approach, we could find a solution  ${\bf A}$, but it would be asymmetric.}, i.e., ${\bf A}={\bf A}^{\top}$.
\end{enumerate}
Consequently, the set of valid adjacency matrices $\ccalA$ is defined as a set of symmetric matrices with no self-loops.
{Problem~\eqref{e:sem_graph_learning} is a constrained convex optimization problem~\cite{boyd2004convex}. We solve it using CVX~\cite{grant2014cvx}, an off-the-shelf convex optimization solver based on interior-point methods that ensures convergence to the global optimum for convex problems. Given the moderate dimensionality of the considered datasets ($N=124$ for EMO and $N=145$ for ARAUS), this approach is computationally tractable and numerically stable.}
{Nevertheless, first-order methods such as proximal gradient descent (or its accelerated variant, FISTA) can be directly applied to~\eqref{e:sem_graph_learning}. In that case, each iteration requires computing $\bbA\bbX$, leading to a per-iteration computational cost on the order of $\mathcal{O}(N^2 M)$. The corresponding convergence rate is $\mathcal{O}(k^{-1})$ for proximal gradient and $\mathcal{O}(k^{-2})$ for FISTA~\cite{beck2017first,navarro2024fair}.}

Solving the graph learning problem in~\eqref{e:sem_graph_learning} yields an estimate $\hbA_\lambda$ that encodes how each feature can be predicted from its neighbors, including those related to perceived emotions, such as arousal and valence. 
Although these two outputs are often the main focus of SER, the proposed model allows any feature to be predicted from other features in the graph, providing a general and interpretable framework for analyzing inter-feature dependencies.  
{Under the SEM assumption, strongly correlated features typically lead to smaller reconstruction errors. In settings where many nodes exhibit high correlation, multiple sparse adjacency matrices may attain comparable reconstruction performance. 
While this does not affect the identification of relevant feature subsets in our framework, alternative Lasso-type formulations designed for correlated features can be employed in such scenarios~\cite{zou2005elastic}.}
Unlike alternative methods, we explicitly constrain $\hbA_\lambda$ to be symmetric, reflecting the assumption of an undirected feature graph in which the influences between pairs of features are bidirectional.  
Moreover, prior studies suggest that arousal and valence are statistically independent or only weakly related, with large variations depending on the context~\cite{kuppens2013relation}. 
This property can be naturally incorporated into the model via additional structural constraints, and its implications are examined in detail in the numerical evaluation in Section~\ref{s:experiments}.
{In practice, numerical optimization may produce very small nonzero entries in $\hbA_\lambda$. To enforce sparsity, coefficients with magnitude below a small threshold $\delta>0$ are set to zero.}

Finally, the regularization parameter $\lambda$ plays a central role in the feature selection process because it directly controls the sparsity of the learned graph.
Larger values of $\lambda$ promote sparser structures, often yielding more interpretable relationships at the cost of a higher reconstruction error, whereas smaller values produce denser graphs with greater fidelity to the data.  
Thus, the selection $\lambda$ involves a trade-off between interpretability and prediction accuracy.
Developing a strategy for choosing $\lambda$ is a non-trivial task, which we address in the next section.

\section{Choosing $\lambda$ by automatic elbow detectors}\label{s:elbow_detector}
{This section introduces a novel method to determine the value of $\lambda$. We first present the some preliminaries, then describe the generalized UAED technique, and finally apply it to the graph learning framework for selecting $\lambda$.}

\begin{figure}[tb]   \centerline{
    \includegraphics[width=0.6\columnwidth]{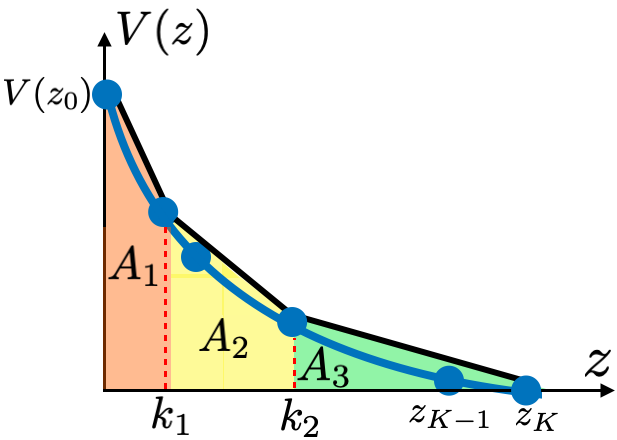}
       }
   \caption{Example of error curve $V(z)$ (blue solid line)  and the construction  of the areas $A_1$, $A_2$, $A_3$, with three straight lines, with $k_1<k_2$. The curve is sampled in $K=6$ points at $z_0$, $z_1$...,  $z_K$,  depicted with blue circles.
   For simplicity, we have consider $z_0=0$ and $V(z_K)=0$.}
   \label{fig:Curva_Inicial}
   \vspace*{-.3cm}
\end{figure}

In several applications, we can construct a {\it non-increasing error curve} (e.g., the mean squared error).
{Formally, denote an error function of interest as}
$
V(z): \mathbb{R} \rightarrow \mathbb{R},
$
{where $z$ can represent a discrete or continuous input variable. Intuitively, $z$ is discrete when representing the number of parameters or the order of a model, and continuous when related to the regularization coefficient in a regularized regression problem.}
A graphical example is shown  in Figure~\ref{fig:Curva_Inicial}.
{To illustrate this setting, consider the case where $\lambda$ is the regularization parameter of a model and define $z = 1/\lambda$.}
In this scenario, as $z\rightarrow \infty$ (i.e., $\lambda \rightarrow 0$), $V(z)$ approaches the minimum achievable error, which typically corresponds to an overfitted solution.
{For instance, if the minimum achievable error is 0, then $V(z)\rightarrow 0$ as $z\rightarrow \infty$.}
However, when using a regularizer, the objective is typically to find a trade-off between data fitting and generalization.
In other words, there exists an optimal value $z^*$.
{Determining and/or defining this optimal value $z^*$ remains an active research problem~\cite{UAEDmorgado2023universal,SICmartino2023spectral,Aho14,Elbow_paper}.}

{In practice, instead of knowing the value of the function $V(z)$ for every $z$, we can only evaluate it at a finite set of $K$ points $z_1< z_2<...<z_K$, sampled (generally non-uniformly) from the domain of $z$.}
Thus, we finally observe $V(z)$ in a finite number of points, 
$$
V(z_1), V(z_2),...V(z_K).
$$
Without loss of generality, we assume that $V(z_K)=0$. 
{Note that the spacing between consecutive points may vary, i.e., $z_{k+1}-z_k$ can depend on $k$, and in general $|z_{k+1}-z_k| \neq 1$.}
{Our goal is to determine an interval of values $[k_1^*,k_2^*]$ that likely contains the optimal value $z^*$ and, more generally, represents a set of practically suitable choices for $z$. To this end,}
We generalize the universal automatic elbow detector (UAED) in~\cite{UAEDmorgado2023universal} in three directions:

\begin{itemize}

    \item {The method accounts for non-uniform sampling in the $z$-domain, i.e., distances $|z_{k+1}-z_k|$ may vary with $k$.}
    
    \item The new technique can handle continuous values of the inputs $z$ in $V(z)$.
    {This means that $z$ can be either a discrete or a continuous variable, with the discrete case being a particular instance.}
    
    \item {The new algorithm returns an interval $[k_1^*,k_2^*]$ rather than a single elbow point.} This is important because, in many applications, there exists a range of values that yield similar performance, allowing the user to choose the most suitable value in this range. Moreover, this procedure also provides information on the uncertainty in the system.

\end{itemize}

\subsection{Interval solution by the generalized UAED (G-UAED)}\label{section2}
{In the following, we introduce the generalized UAED (G-UAED), a novel technique for obtaining an uncertainty interval in the context of elbow detection.}
{Let}
$$
{k_1, k_2 \in \{z_0,z_1,...,z_K\}, \quad  \mbox{ and } k_1\leq k_2,}
$$
{so that the two points defining the interval solution are denoted by $[k_1^*,k_2^*]$.}
Note that if $V(z)$ has a significant drop at the beginning, for instance, reaching zero already at the first point $z_1$ (i.e., $V(z_1)=0$), the elbow will be located at $z_1$.
{Conversely, if $V(z)$ decreases approximately linearly and reaches zero only at the last point $z_K$, then there is no clear elbow and the only possible choice is $z_K$.}
In general, a smaller area under $\mathrm{V}(\mathrm{z})$ implies the elbow is shifted toward the origin along the z axis, i.e., toward $z=0$.
{Therefore, our goal is to approximate the area under $V(z)$ and derive a rule to determine the elbow point.}

{In this context}, we build a piecewise-linear approximation of the curve $V(z)$.
{The underlying idea is to approximate $V(z)$ using three straight-line segments: the first one connecting the points $(z_0,V(z_0))$ and $(k_1,V(k_1))$, the second connecting $(k_1,V(k_1))$ and $(k_2,V(k_2))$, and the third connecting $(k_2,V(k_2))$ and $(z_K,0)$ (recall that we have assumed $V(z_K)=0$).}
Thus, inspired by the concept of {\it the  maximum area under the curve in ROC},  the goal is to minimize the area under this piece-linear approximation of the curve $V(z)$.
The total area under this approximation is the sum of the two trapezoidal areas  ($A_1$ and $A_2$) and a triangular area ($A_3$), as depicted in Figure \ref{fig:Curva_Inicial}. {This areas are given by}
\begin{align} \label{A_again}
A_1 &= \frac{(V(z_1) +V(k_1)) (k_1-z_0)}{2}, \nonumber \\
A_2 &=  \frac{(V(k_1) +V(k_2)) (k_2-k_1)}{2}, \quad \\
 A_3 &=\frac{V(k_2) (z_K-k_2)}{2}.  \nonumber
\end{align}
The final solution is given by
\begin{align}\label{OptEq}
 [k_1^*,k_2^*] =\arg\min_{k_1,k_2}\{A_1 + A_2 + A_3\},  \quad \text{s.t.} \; k_1<k_2,
\end{align}
where $A_1$, $A_2$, and $A_3$ are defined in~\eqref{A_again}.
{It can be shown that the optimal interval $[k_1^*,k_2^*]$ contains the elbow given by the standard UAED in~\cite{UAEDmorgado2023universal}.}
{Although the optimization problem in~\eqref{OptEq} is generally not convex, the search space is finite and tractable.}
{Indeed, the candidate solutions correspond to all pairs $(k_1,k_2)$ with $k_1<k_2$, yielding a number of possible intervals on the order of $\ccalO(K^2)$.}
{Therefore, when $K$ is moderate, the optimal solution can be obtained via exhaustive search.
Finally, the optimal value $k^*$ is obtained following~\cite{UAEDmorgado2023universal}.}

{\subsection{G-UAED for sparsity regularization parameter selection}}\label{section3}
{Our feature selection procedure combines the graph learning problem in \eqref{e:sem_graph_learning} with the G-UAED technique described in the previous subsection. This process is summarized in Algorithm~\ref{alg:sem_guaed} and is detailed below.

\algrenewcommand\algorithmicrequire{\textbf{Input:}}
\algrenewcommand\algorithmicensure{\textbf{Output:}}

\begin{algorithm}[t]
\caption{Feature Selection via graph SEM and G-UAED.}\label{alg:sem_guaed}

\begin{algorithmic}[1]
\Require Data $\mathbf{X} \! \in \! \mathbb{R}^{N \times M}$; regularization parameters $\{\lambda_k\}_{k=0}^K$
\Ensure $\lambda^*$, $[\lambda_1^*,\lambda_2^*]$, $\widehat{\mathbf{A}}_{\lambda^*}$, Ranking

\For{$k = 0$ to $K$}
    \State Estimate $\widehat{\mathbf{A}}_{\lambda_k}$ by solving \eqref{e:sem_graph_learning} for $\lambda_k$
    \State Set $\mbox{N-MSE}(\lambda_k) =
        \| \mathbf{X} - \widehat{\mathbf{A}}_{\lambda_k} \mathbf{X} \|_F^2 / \|\mathbf{X}\|_F^2$
    \State Set $\epsilon_k = \| \widehat{\mathbf{A}}_{\lambda_k} \|_0$
\EndFor

\State Get $[\lambda^*_1,\lambda^*_2], \lambda^*$ solving
    $\text{G-UAED} \left(\{ \epsilon_k, \mbox{N-MSE}(\lambda_k)\}_{k=0}^K \right)$ as in (5)
\State Set $\widehat{\mathbf{A}}_{\lambda^*}$ to the estimate associated with $\lambda^*$
\State $\text{Importance} \gets \text{Feature Importance}(\widehat{\mathbf{A}}_{\lambda^*})$
\State $\text{Ranking} \gets \text{Sort Descending}(\text{Importance})$

\Return $\lambda^*$, $[\lambda_1^*,\lambda_2^*]$, $\widehat{\mathbf{A}}_{\lambda^*}$, $\text{Ranking}$
\end{algorithmic}
\end{algorithm}

Recall that we aim to learn the graph structure by solving \eqref{e:sem_graph_learning}, which involves a trade-off between promoting sparsity and minimizing the reconstruction error.
If $\lambda=0$, we obtain the matrix $\hbA_\lambda$ that best fits the data (i.e., minimizing the error $\| \bbX - \hbA_\lambda\bbX \|_F^2$), and the corresponding graph contains the maximum number of edges.
As $\lambda$ increases, the reconstruction error $\| \bbX - \hbA_\lambda\bbX \|_F^2$ also increases, while the matrix $\hbA_\lambda$ becomes sparser, and the corresponding graph contains fewer edges.
In the limit, when $\lambda$ becomes sufficiently large, say $\lambda \geq \lambda_{\texttt{max}}$, the matrix $\hbA_\lambda$ becomes the null matrix.

Under this setting, the reconstruction error is determined by the regularization parameter $\lambda$.
In particular, let the normalized Frobenius error be defined as
\begin{equation}\label{NormErrEq}
    \mbox{N-MSE}(\lambda)
    = \frac{\|{\bf X}-\hbA_\lambda {\bf X}\|^2} {\|{\bf X}\|^2}.
\end{equation}
With a slight abuse of notation, we set $z=\lambda^{-1}$ and use $\mbox{N-MSE}(\lambda)$ as the error function, yielding
\begin{equation}
    V(z)=\mbox{N-MSE}(\lambda^{-1}).
\end{equation}
This change of variable is convenient because the reconstruction error is typically non-decreasing with respect to $\lambda$, whereas the G-UAED operates on non-increasing error curves. Hence, defining $z=\lambda^{-1}$ ensures the required monotonic behavior of $V(z)$.

Then, to determine the optimal value of $\lambda$, and hence the optimal number of edges, we define a grid of increasing values $\lambda_0 < \lambda_1 < \cdots < \lambda_K$, ranging from $\lambda_0=0$ to $\lambda_K=\lambda_{\texttt{max}}$.
For each $\lambda_k$, we solve the optimization problem in \eqref{e:sem_graph_learning}, obtaining the estimate $\hbA_{\lambda_k}$, computing the error $\mbox{N-MSE}(\lambda_k)$, and evaluating the number of edges through the $\ell_0$ norm $\epsilon_k := \| \hbA_{\lambda_k} \|_0$.
This procedure yields a non-decreasing sequence of reconstruction errors, i.e., 
$\mbox{N-MSE}(\lambda_{k+1}) \geq \mbox{N-MSE}(\lambda_{k})$ 
(Note that the order of $\lambda$ is reversed to match the natural ordering of $z$, since $z=\lambda^{-1}$).
Given the sequence of values $V(\lambda_k^{-1})$, we employ the G-UAED to determine the optimal interval of $\lambda$ values.
Note that different error functions can also be accommodated, with $V(z)=\log(\mbox{N-MSE}(z))$ being an alternative similar to other information criteria.
Solving the optimization problem in \eqref{OptEq} yields an optimal choice of $z^*=1/\lambda^*$ together with a confidence interval that quantifies the uncertainty in the selection.
Once $\hbA_{\lambda^*}$ is obtained, we derive a feature ranking for each node (i.e., each SER variable) by ordering the remaining variables according to the magnitudes of the corresponding edge weights in the adjacency matrix.

While we could directly employ $z=\lambda^{-1}$, in practice, we consider $z$ as the number of edges $\epsilon_k$ in the graph, which is itself a function of $\lambda$.
This choice was made for two reasons. First, because the goal is feature selection, we are interested in identifying the smallest number of edges that still allows an accurate reconstruction.
Second, this choice tends to produce smoother and more convex error curves, which facilitates the application of the G-UAED.

Note that the optimal selection of $\lambda$, which balances model fitting, generalization, and other desirable properties, remains an active area of research~\cite{FreijeiroLASSO,TibshiraniLASSO}. 
In practice, the literature suggests that there often exists a range of acceptable $\lambda$ values yielding similar performance, rather than a unique optimal value. 
This behavior is particularly well documented for LASSO-type regularization in variable selection problems, where multiple values of $\lambda$ may provide satisfactory solutions, as illustrated in the next section.}

\section{Numerical experiments}\label{s:experiments}
{In this section, we apply the proposed method to the EMO dataset to analyze the relationships among its acoustic features and emotional dimensions. We also report a brief synthetic experiment to validate G-UAED as a regularization-selection strategy and analyze the ARAUS dataset to illustrate the broader applicability of the framework. The code used to reproduce all experiments is publicly available on GitHub\footnote{\url{https://github.com/reysam93/acoustic\_nti}}.}


\begin{figure*}[tb]
      \centerline{
      \subfigure[\label{gammaval1} Scenario {C1}.]{\includegraphics[width=0.7\columnwidth]{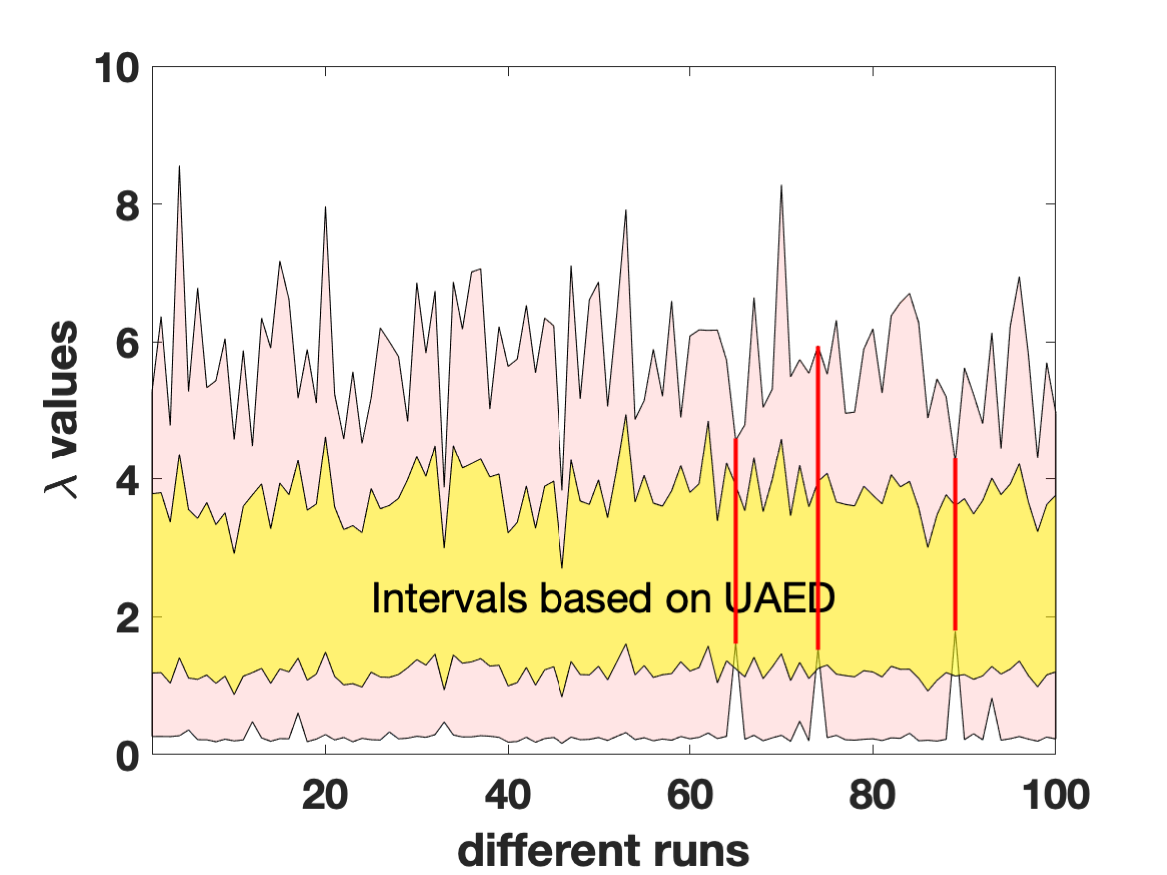}}
            \subfigure[\label{gammaval2} Scenario {C2}.]{\includegraphics[width=0.7\columnwidth]{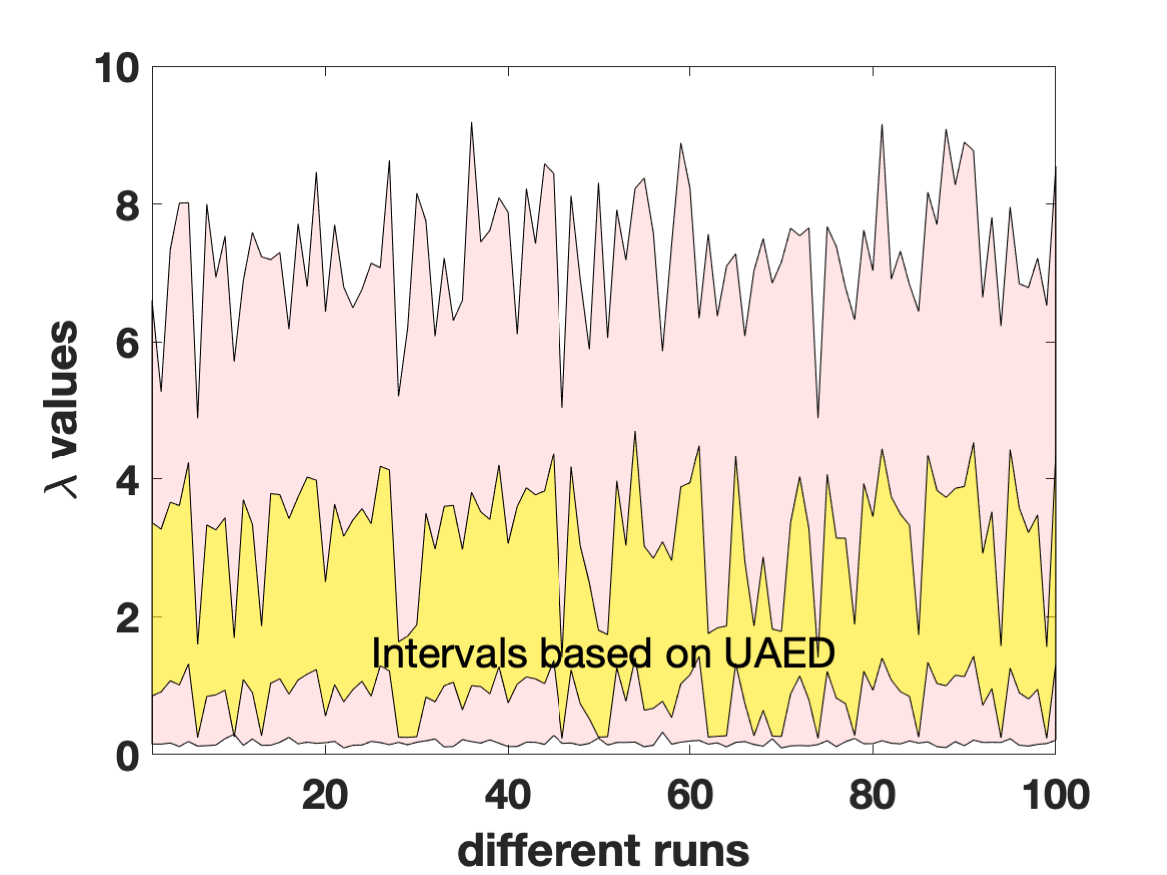}}
  }
    \caption{The ground-truth intervals $[\lambda_1^*,\lambda_2^*]$ in 100 different realizations are depicted with pink shaded areas. Within these intervals, the $\lambda$ values yield exactly $10$ correctly located zeros in ${\bm \beta}^*_\lambda$ in {configuration C1}, {and $20$ correctly located zeros in ${\bm \beta}^*_\lambda$ in configuration C2}. The intervals provided by G-UAED are shown by the yellow shaded areas in {\bf (a)} for {configuration C1} and in {\bf (b)} for {configuration C2}. {In configuration C1, only 3 runs (highlighted in red) contain a small portion of the G-UAED interval outside the set of suitable values. In configuration C2, the G-UAED intervals always contain suitable values.}}
   \label{fig:Picasso_comeme_el_casso3}
   \vspace*{-.3cm}
\end{figure*}

\subsection{{Validating G-UAED for regularization parameter selection}}\label{LassoEx}

{We now assess the ability of G-UAED to suggest a suitable interval of $\lambda$ values.
Consider a dataset $\{{\bf x}_i,y_i\}_{i=1}^N$, where ${\bf x}_i=[x_{i,1},...,x_{i,R}]^{\top}$ denotes the input vectors collected in the matrix ${\bf X}=[{\bf x}_1,...,{\bf x}_N] \in \reals^{R\times N}$,
and ${\bf y}=[y_1,...,y_N]^{\top}$ is the output vector.}
We also define the vector of unknown parameters ${\bm \beta}=[\beta_1,...,\beta_R]^{\top}$, and assume the model
$$
{\bf y}={\bm \beta}^\top{\bf X}+{\bm \epsilon},
$$
where ${\bm \epsilon}$ is a noise perturbation that we consider Gaussian with zero mean and variance $\sigma^2=1$. We set $R=30$ and generate artificially $N=200$ data pairs following this model by considering
$x_{i,r} \sim \mathcal{N}(x|0,100)$ (independently drawn for each $i$, $r$ in each run).
{We consider two configurations}.
In the first one, we assume the vector
$$
\mbox{{\bf {C1}:}}\quad{\bm \beta}_{\texttt{true}}=[\underbrace{1, 1, 1, 1,..,1}_{20} \underbrace{0, 0, 0, 0,..,0}_{10}]^{\top},
$$
Hence, the last $10$ features do not affect the outputs $y_i$. In a second {configuration}, we assume the vector
$$
\mbox{{\bf {C2}:}}\quad{\bm \beta}_{\texttt{true}}=[\underbrace{1, 1, 1, 1,..,1}_{10} \underbrace{0, 0, 0, 0,..,0}_{20}]^{\top},
$$
i.e., we have more zeros (the $20$ last positions).

{After generating the data, we apply LASSO~\cite{FreijeiroLASSO,TibshiraniLASSO} to obtain an estimate of the coefficient vector ${\bm \beta}_{\lambda}^{*}$.
Specifically, we solve the following $\ell_1$-regularized least-squares problem}
\begin{align}\label{RegForm}
{\bm \beta}^*_\lambda=\arg\min_{\beta}\Big[||{\bf y}-{\bm \beta}^\top{\bf X}||^2 +\lambda ||{\bm \beta}||_1\Big], \quad \lambda>0.
\end{align}

{We use a fine grid of $\lambda$ values. As $\lambda$ grows, the number of zeros in ${\bm \beta}^*_\lambda$ increases, up to a maximum value of $\lambda$ such that ${\bm \beta}^*_\lambda$ is the null vector.} We denote as
$$
    \mbox{MSE}(\lambda)=\|{\bf y}-({\bm \beta}_\lambda^*)^\top{\bf X}\|_2^2.
$$
Clearly, with $\lambda=0$, we obtain the smallest MSE, and if $0<\lambda_1<\lambda_2$ we have $\mbox{MSE}(\lambda_1) < \mbox{MSE}(\lambda_2)$, i.e., $\mbox{MSE}(0) \leq \mbox{MSE}(\lambda_1) \leq \mbox{MSE}(\lambda_2)$.
The optimal selection of $\lambda$, which balances model fitting, generalization, and other desirable properties, remains an active area of research \cite{FreijeiroLASSO,TibshiraniLASSO}.
{In practice, there is broad agreement that a range of $\lambda$ values may be acceptable, rather than a single optimal choice. This is particularly clear for LASSO-based variable selection, where multiple values of $\lambda$ may yield the correct sparsity pattern in ${\bm \beta}_\lambda^*$.}
Note that, if we set $z=1/\lambda$ and define
\begin{align}
    V(z)=\mbox{MSE}(z)=\mbox{MSE}(1/\lambda),
\end{align}
we have a non-increasing error function of $z$.
{Therefore, we can apply G-UAED to provide an optimal choice of $z^*=1/\lambda^*$, together with an interval solution $[k_1^*,k_2^*]$ such that $z^*\in [k_1^*,k_2^*]$, thereby quantifying the uncertainty in the selection.
However, instead of defining the variable as $z=1/\lambda$, we consider $z$ to be the number of non-zero coefficients in the model for a given $\lambda$. When $\lambda=0$, this number is maximal, whereas for $\lambda=\lambda_{\max}$ it is zero.} The motivation behind this choice is that it tends to produce smoother and more convex curves, thereby facilitating the application of the G-UAED~\cite{UAEDmorgado2023universal}.

{For each configuration, let $\lambda_1^*$ denote the smallest value of $\lambda$ and $\lambda_2^*$ the largest value of $\lambda$ such that ${\bm \beta}_\lambda^*$ recovers the correct sparsity pattern, i.e., the true zero entries are correctly identified.}
Hence, reasonable choices of $\lambda$ for our problem {lie in the interval} $[\lambda_1^*,\lambda_2^*]$. For each realization of the data, the ground-truth interval $[\lambda_1^*,\lambda_2^*]$ is shown in Figure~\ref{fig:Picasso_comeme_el_casso3} as a pink shaded area.
We apply the procedures in Section \ref{s:elbow_detector} to find the interval solutions based on the G-UAED.
{As shown in Fig.~\ref{fig:Picasso_comeme_el_casso3}, the intervals suggested by G-UAED are fully contained in the corresponding ground-truth intervals $[\lambda_1^*,\lambda_2^*]$ in all but $3$ out of $100$ independent runs.}
{In those three cases, only a small portion at the beginning of the estimated interval lies slightly outside the ground-truth interval. These runs are highlighted in red.}
Hence, the interval solutions provided by G-UAED contain safe and useful $\lambda$ values for use in a LASSO  regression problem.

\subsection{Feature selection on the EMO database (experiments)}
{We now apply the proposed framework to the EMO dataset described in Section~\ref{s:emosoundscape}. Recall that each soundscape is represented by 122 acoustic input features together with the two emotional outputs, arousal and valence, which are jointly collected in the data matrix $\bbX \in \reals^{124 \times 1213}$. Under the SEM assumption introduced in Section~\ref{s:graph_learning}, our goal is to estimate a sparse graph that captures the dependencies among all variables without explicitly separating the inputs from the outputs. To this end, we apply Algorithm~\ref{alg:sem_guaed}, which involves solving the optimization problem in~\eqref{e:sem_graph_learning} for different values of $\lambda$ and then using the G-UAED-based selection strategy described in Section~\ref{s:elbow_detector} to identify suitable sparsity levels.}

In this section, we consider four scenarios:
\begin{itemize}
    \item {\bf Scenario 1.} We allow the connection between the outputs, arousal, and valence (last two rows in ${\bf X}$) and compute the normalized error in~\eqref{NormErrEq} considering all the matrices ${\bf X}$ ({i.e., using the reconstruction of the whole matrix ${\bf X}$}).
    \item {\bf Scenario 2.} We \emph{do not} allow the connection between the outputs and compute the normalized error in~\eqref{NormErrEq} considering all the matrices ${\bf X}$. 
    \item {\bf Scenario 3:} We allow the connection between the outputs, but compute the normalized error in~\eqref{NormErrEq} considering {\it only} the last two rows of ${\bf X}$, so we consider only the reconstruction of the outputs.   
    \item {\bf Scenario 4:} We {\it do not} allow the connection between the outputs, and compute the normalized error~\eqref{NormErrEq} considering {\it only} the last two rows of ${\bf X}$, so we consider only the reconstruction of the outputs.
\end{itemize}

\begin{figure*}[ht]
\centerline{     
             \subfigure[{\bf Scenario 1.}]{\includegraphics[width=0.75\columnwidth]{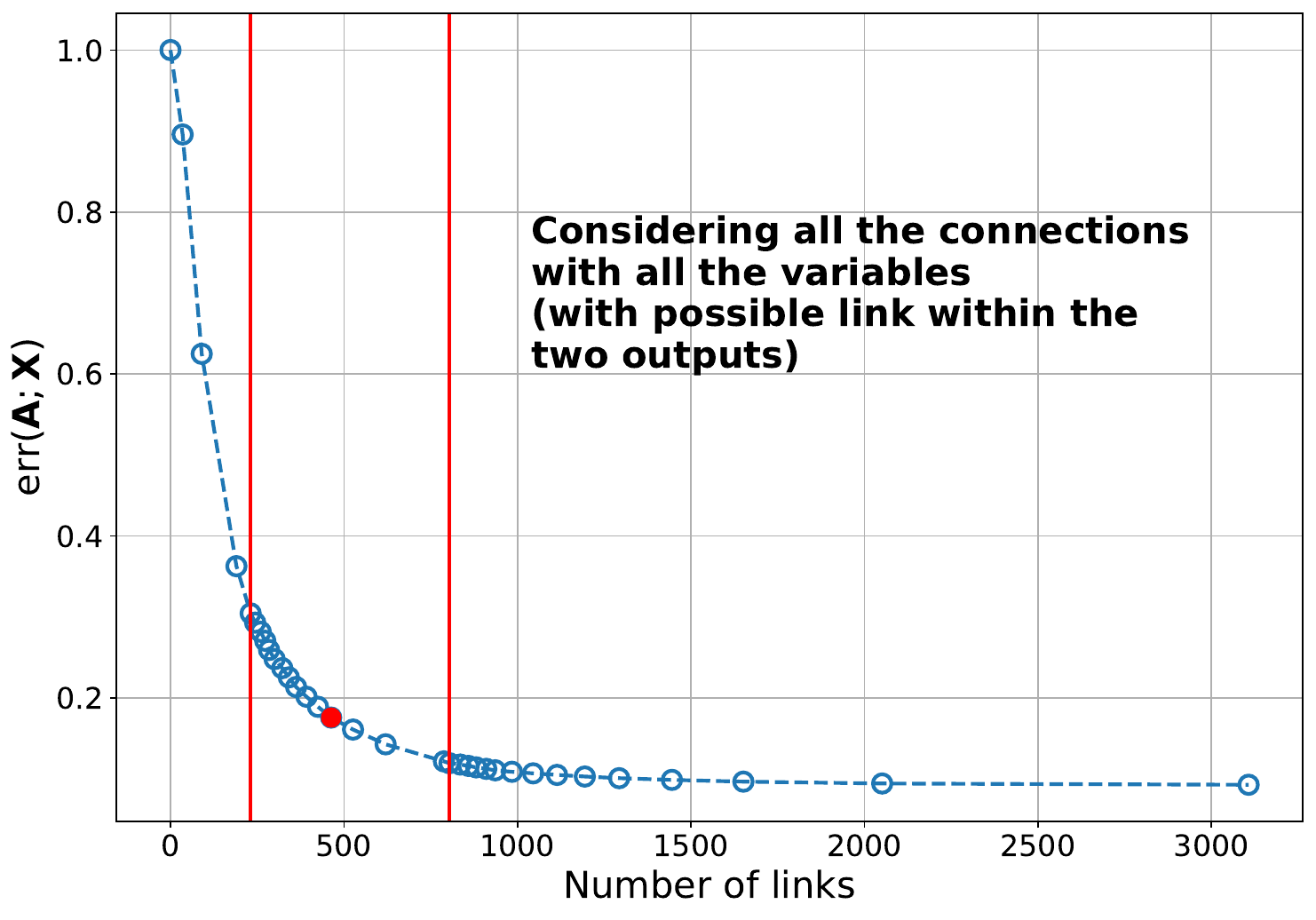}}
               \subfigure[{\bf Scenario 2.}]{\includegraphics[width=0.75\columnwidth]{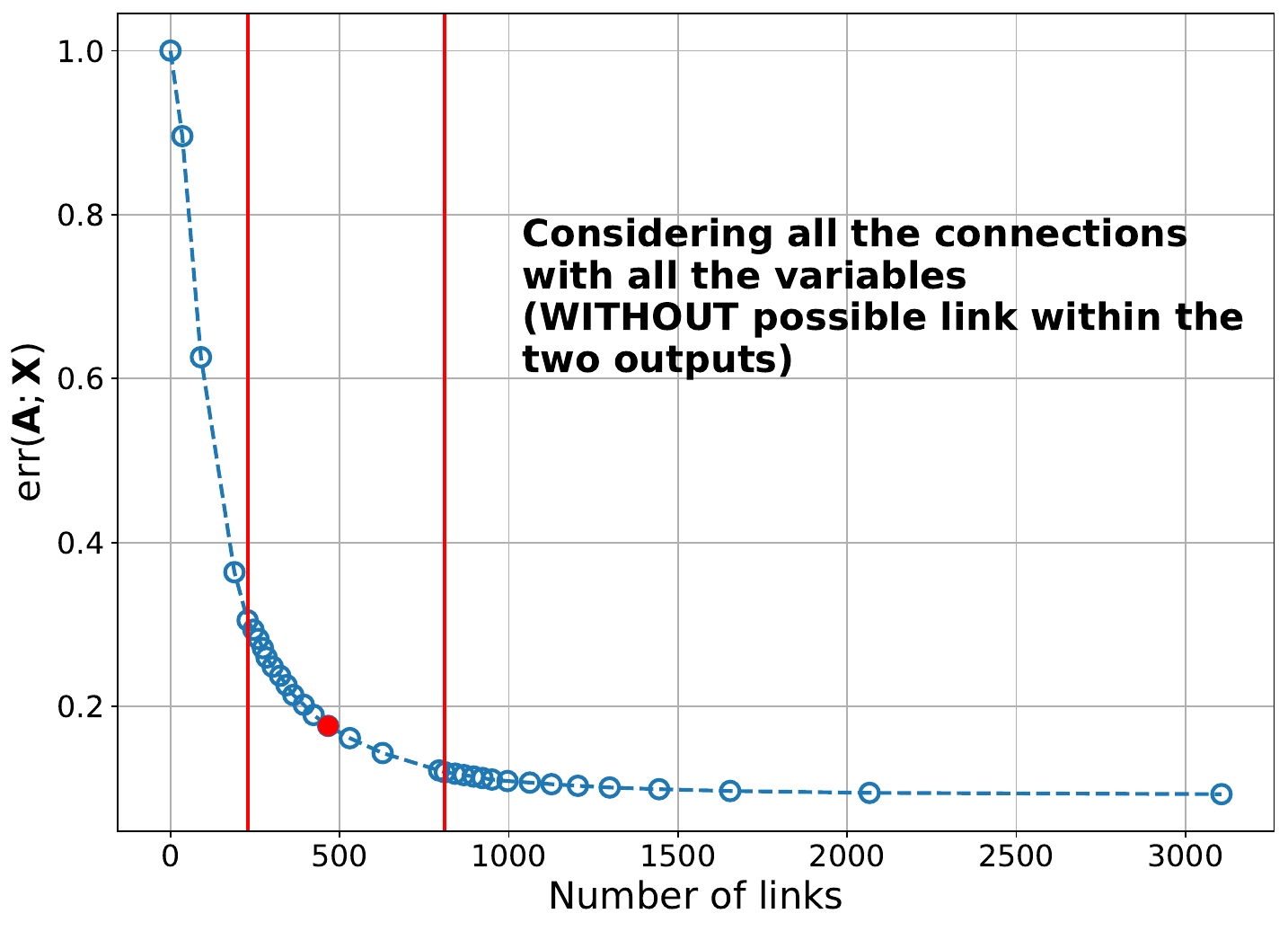}}
  }
  \vspace*{-0.35cm}
      \centerline{
      \subfigure[{\bf Scenario 3.}]{\includegraphics[width=0.75\columnwidth]{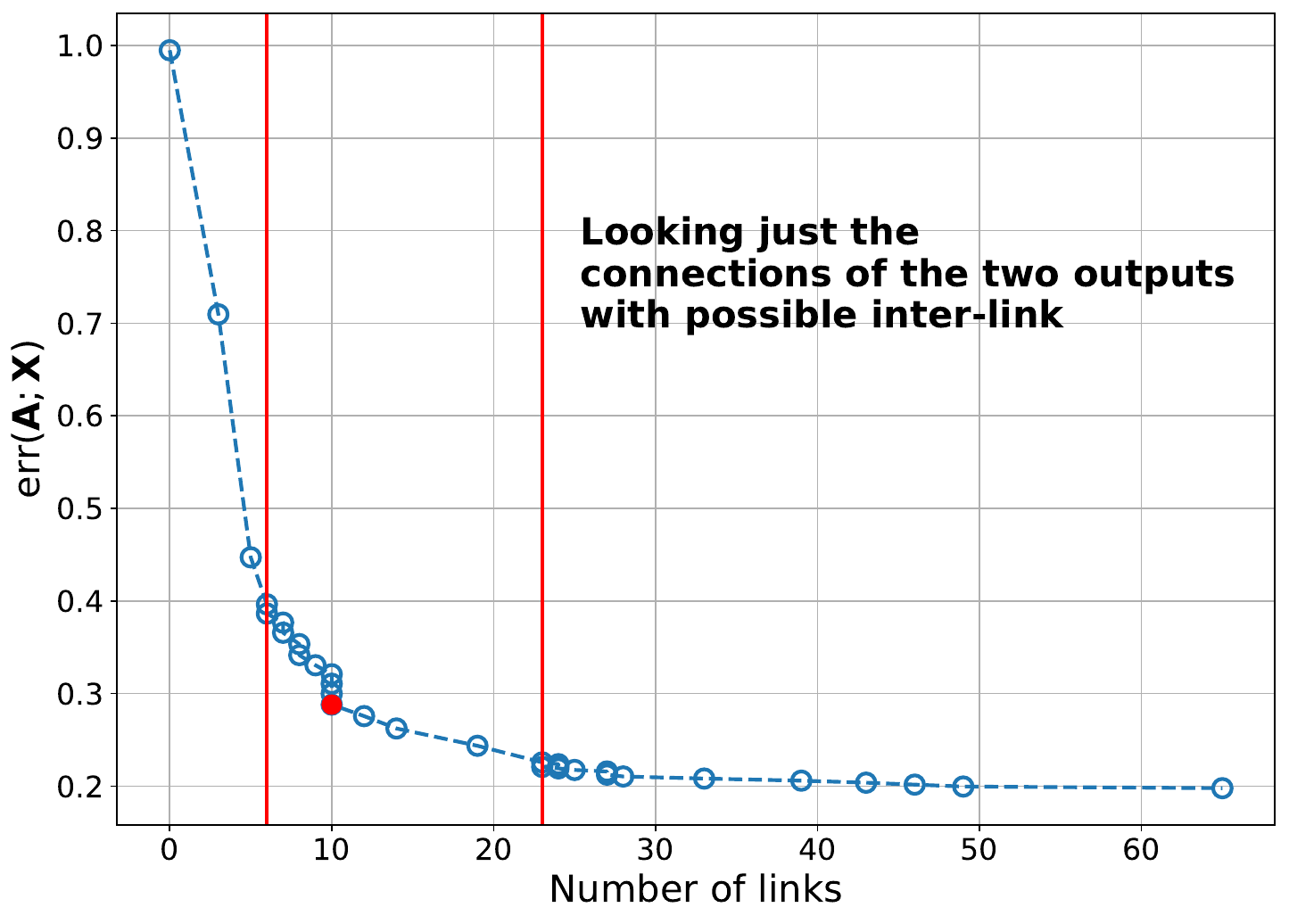}}
         \subfigure[{\bf Scenario 4.}]{\includegraphics[width=0.75\columnwidth]{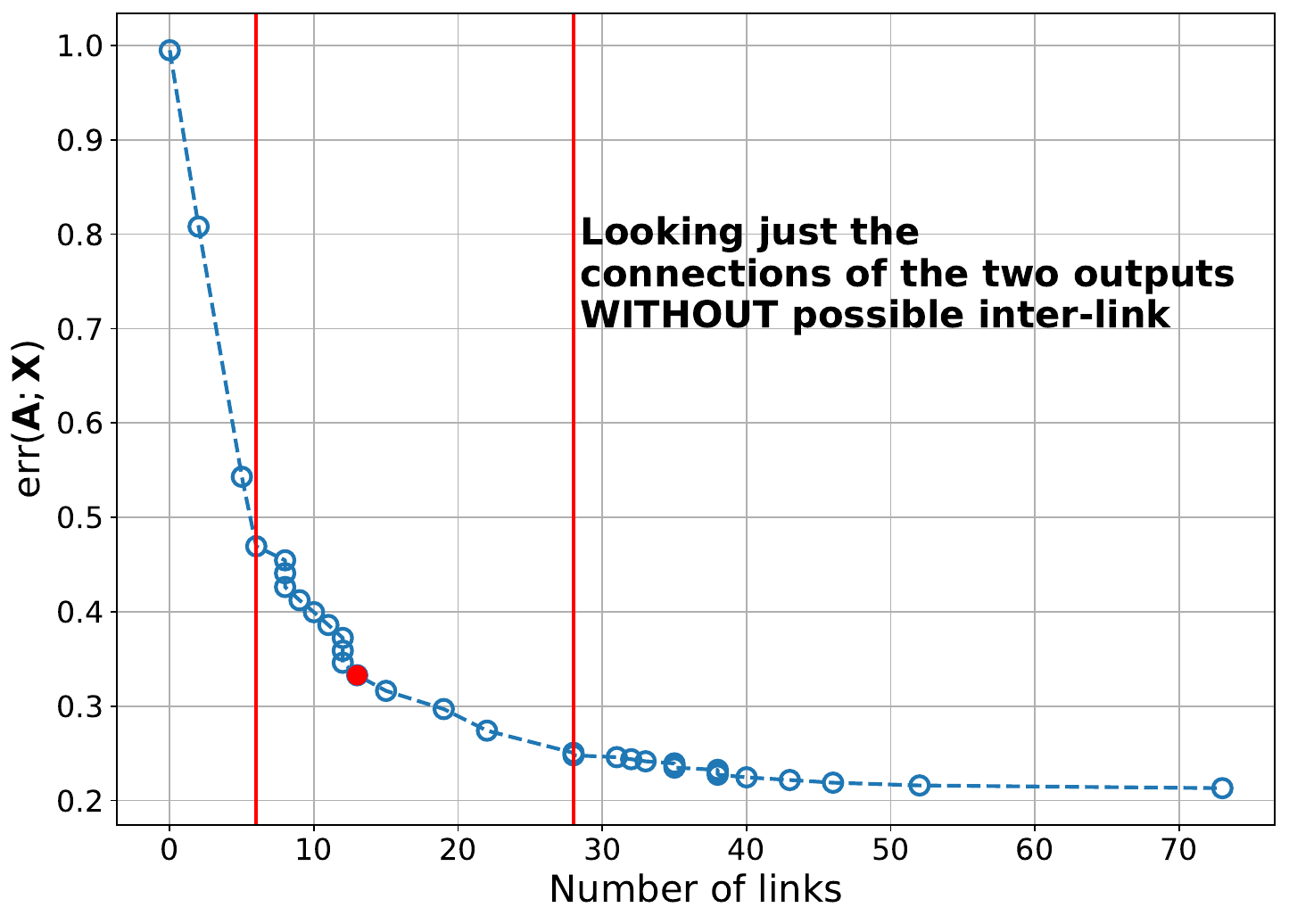}}
         }     
   \caption{The normalized error as function of the number of links in the graph, in the 4 different scenarios. Note that the points with zero links correspond to $\lambda_{\texttt{max}}$. The red points and red lines in each figure correspond to the elbows and intervals obtained by applying the G-UAED.} \label{fig:Picasso_comeme_el_casso3_2}
   \vspace*{-.5cm}
\end{figure*}

The curves showing the normalized error versus the number of links in the graph are depicted in Figure~\ref{fig:Picasso_comeme_el_casso3_2}, for the four scenarios, {where the normalized error~\eqref{NormErrEq} is labeled as $\mbox{N-MSE}(\lambda)=\texttt{err}(\hbA_\lambda;{\bf X})$.}
Clearly, the normalized error decreases as the number of edges increases.
Recall that each point in the figures corresponds to a value of $\lambda$. As remarked above, as the number of connections increases, the corresponding $\lambda$ decreases. Then, we apply the G-UAED scheme in Section~\ref{s:elbow_detector} to obtain the elbows and estimated intervals, represented by the red points and red lines in Figure \ref{fig:Picasso_comeme_el_casso3_2}.
Table \ref{TableElbow} summarizes these results in terms of the links and values of $\lambda$.
{Recall that Scenarios 3 and 4 consider only the two output rows for computing the error, so that the values of the links are smaller.}

\begin{table}[tb]
\begin{center}
\caption{Elbows and intervals (obtained by G-UAED) in  terms of number of links and values of $\lambda$.} \label{TableElbow}
{
\begin{tabular}{|c||c c || c c|} 
 \hline
 \multirow{2}{*}{{\bf Scenarios}} & \multicolumn{2}{c||}{{\bf Links}}  &  \multicolumn{2}{c|}{$\lambda$} \\ 
   \cline{2-5}
& {\bf Elbow} & {\bf Interval} & {\bf Elbow} & {\bf Interval}  \\   
 \hline
  \hline
 Scenario 1 & 463 & [231, 804] & 128.50 & [472.59, 31.50]\\ 
 Scenario 2   & 466 & [228, 811] &  128.50 & [472.59, 31.50]\\ 
 \hline
 Scenario 3   & 10 & [6, 23] & 159.78 & [472.59, 34.65] \\ 
 Scenario 4   & 13 & [6, 28] & 159.78 & [472.59, 34.65]\\ 
 \hline
\end{tabular}
}
\vspace*{-0.3cm}
\end{center}
\end{table}

Looking at Table~\ref{TableElbow},  we can make some interesting observations. {First,} the values of $\lambda$ are identical in Scenarios~1--2 and in Scenarios~3--4. {Moreover, the resulting values are very similar across all four scenarios, with the upper extreme of the interval equal to $472.59$ in every case.}
{Second,} Scenarios~1 and~3 present slightly smaller elbow points in terms of the number of links. This is due to the fact that Scenarios~1--3 allow a direct connection between the output variables, whereas Scenarios~2--4 {explicitly prevent it. This observation suggests an interesting result:} the outputs arousal and valence present a relevant partial correlation/connection. This confirms some hypotheses suggested after the analyses done in previous works \cite{OurPaperSound,san2024variable}, which appear to contrast with theories of psychological phenomena, i.e., challenging the common SER assumptions.

\begin{figure*}[!t]
	\centering
	\begin{subfigure}[Scenario 3: $\lambda=159.78$, with links between the outputs. \label{FigSc3}]{\includegraphics[width=\columnwidth]{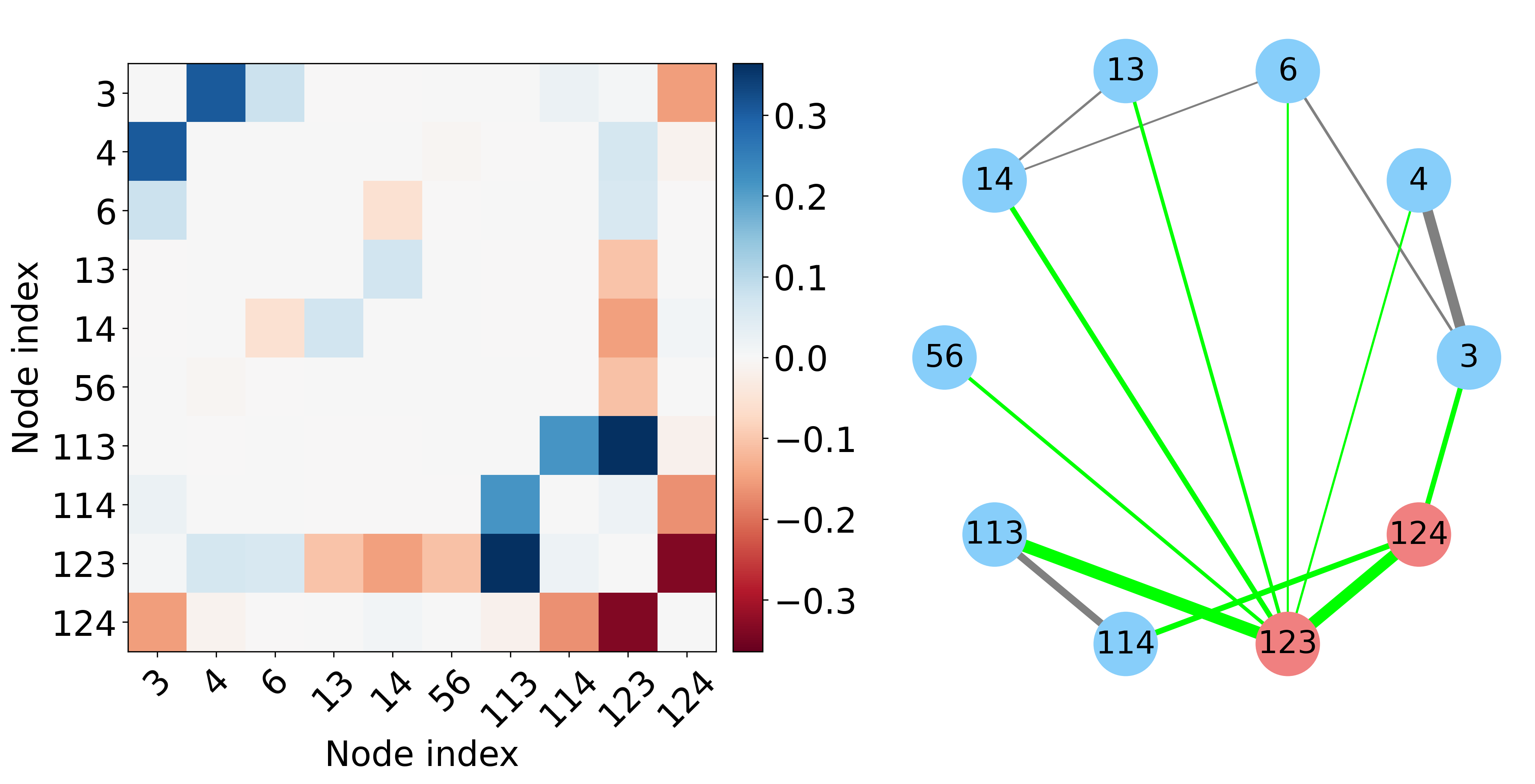}}
	\end{subfigure}
	\begin{subfigure}[With inter-link but $\lambda=300$.]{\includegraphics[width=\columnwidth]{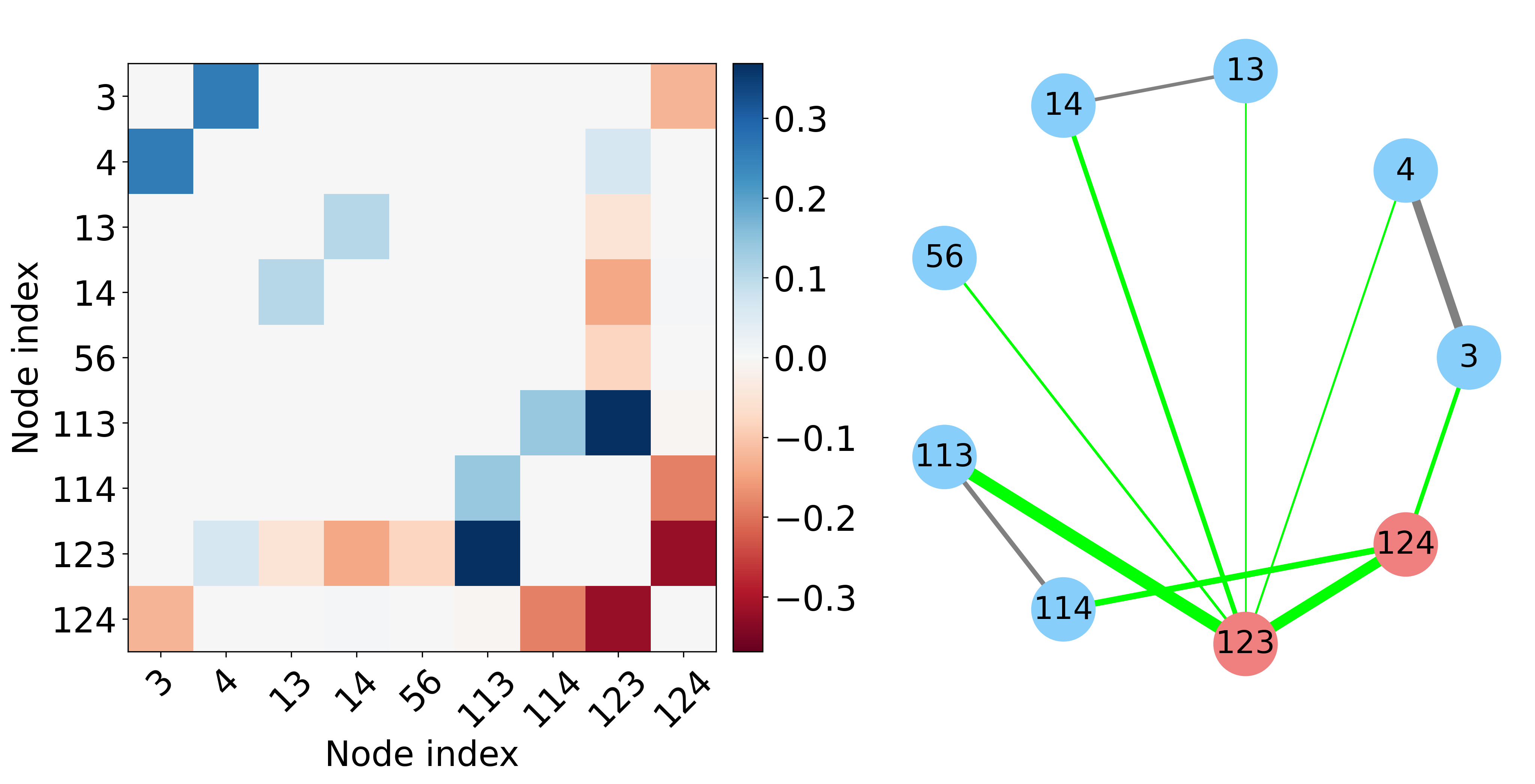}}
	\end{subfigure}
    \begin{subfigure}[With inter-link but $\lambda=360$.]{\includegraphics[width=\columnwidth]{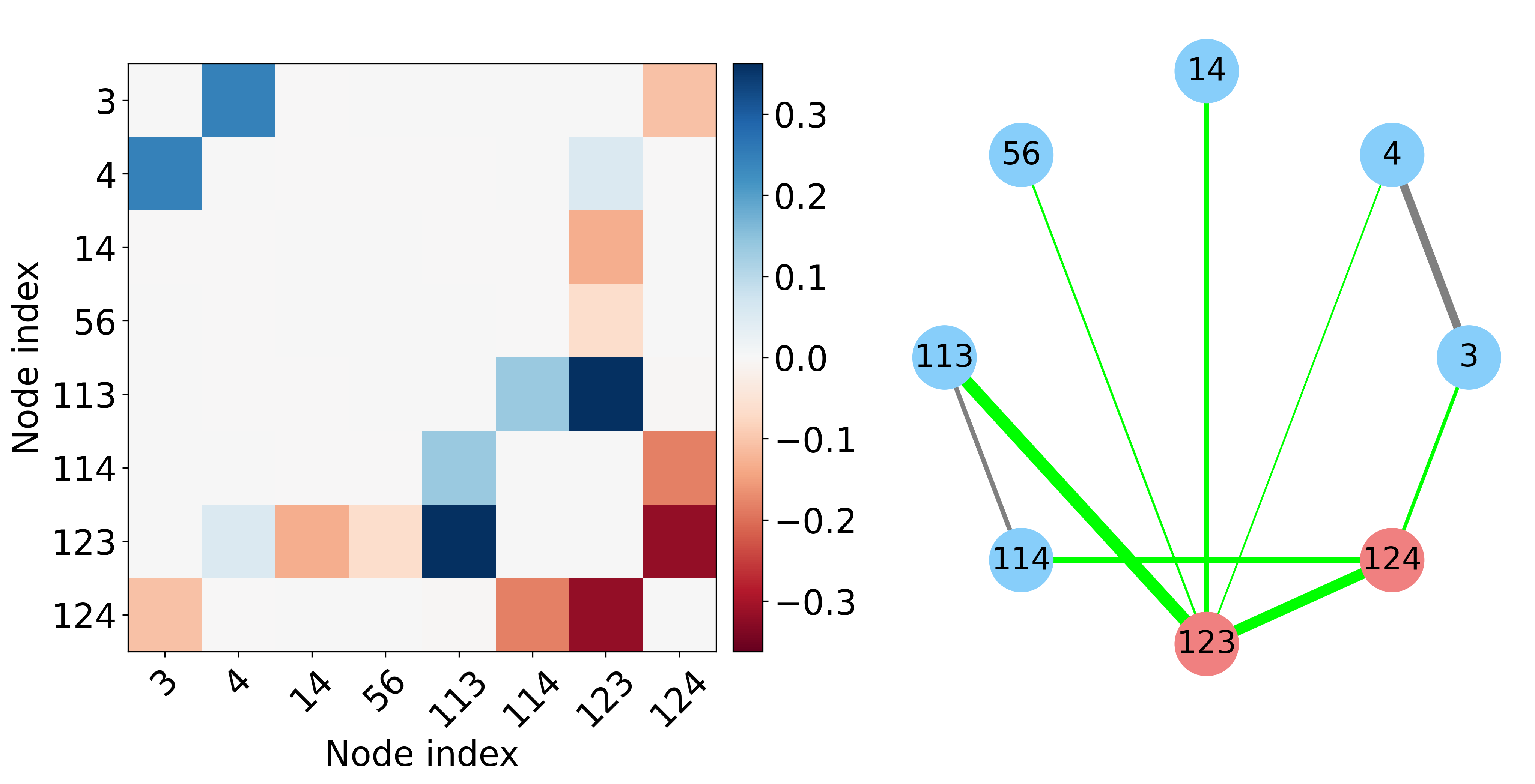}}
	\end{subfigure}
    \begin{subfigure}[With inter-link but $\lambda=472.59$.]{\includegraphics[width=\columnwidth]{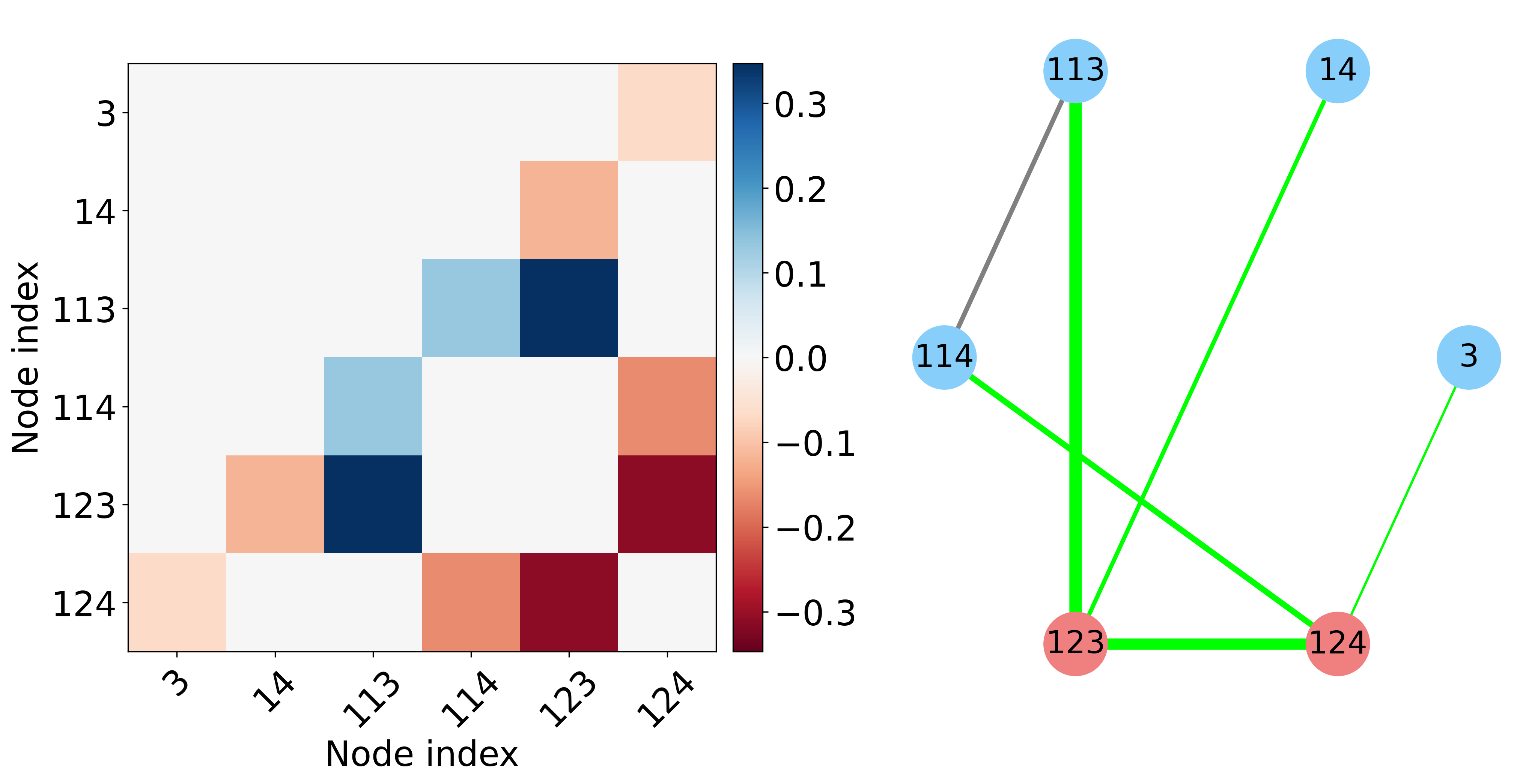}}
	\end{subfigure}
	\caption{Graph representation and associated adjacency matrix of edges involving the output variables (valence and arousal). Connectivity between output variables is allowed. (a) Using the elbow value $\lambda=159.78$; (b) $\lambda=300$; (c) $\lambda=360$; (d) Using the largest value of $\lambda=472.59$ within our interval.}
    \label{f:graphs_with_interlink}
\end{figure*}

\begin{figure*}[!t]
	\centering
	\begin{subfigure}[Scenario 4: $\lambda=159.78$, without links between the outputs. \label{FigSc4}]{\includegraphics[width=\columnwidth]{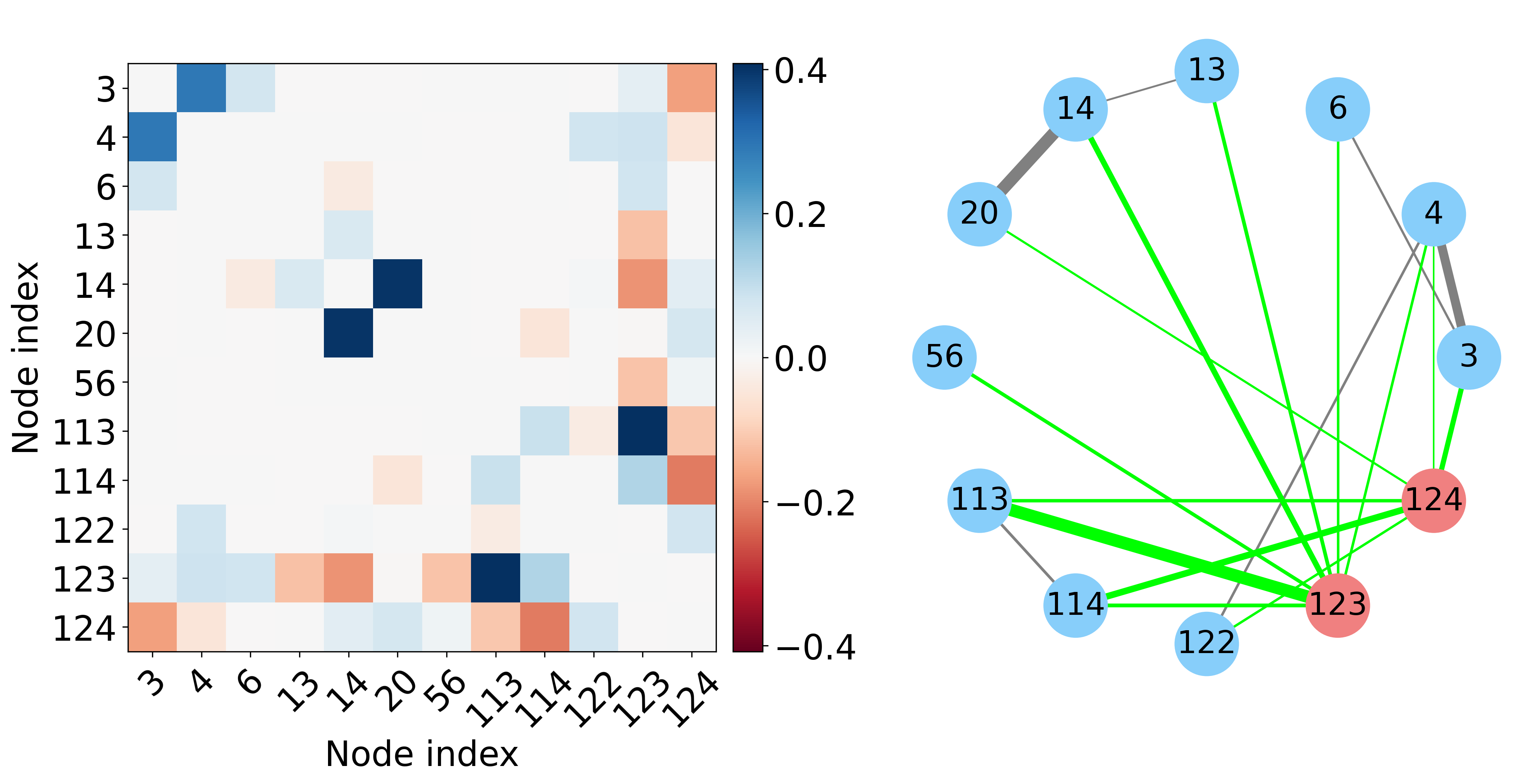}}
	\end{subfigure}
	\begin{subfigure}[Without inter-link but $\lambda=300$.]{\includegraphics[width=\columnwidth]{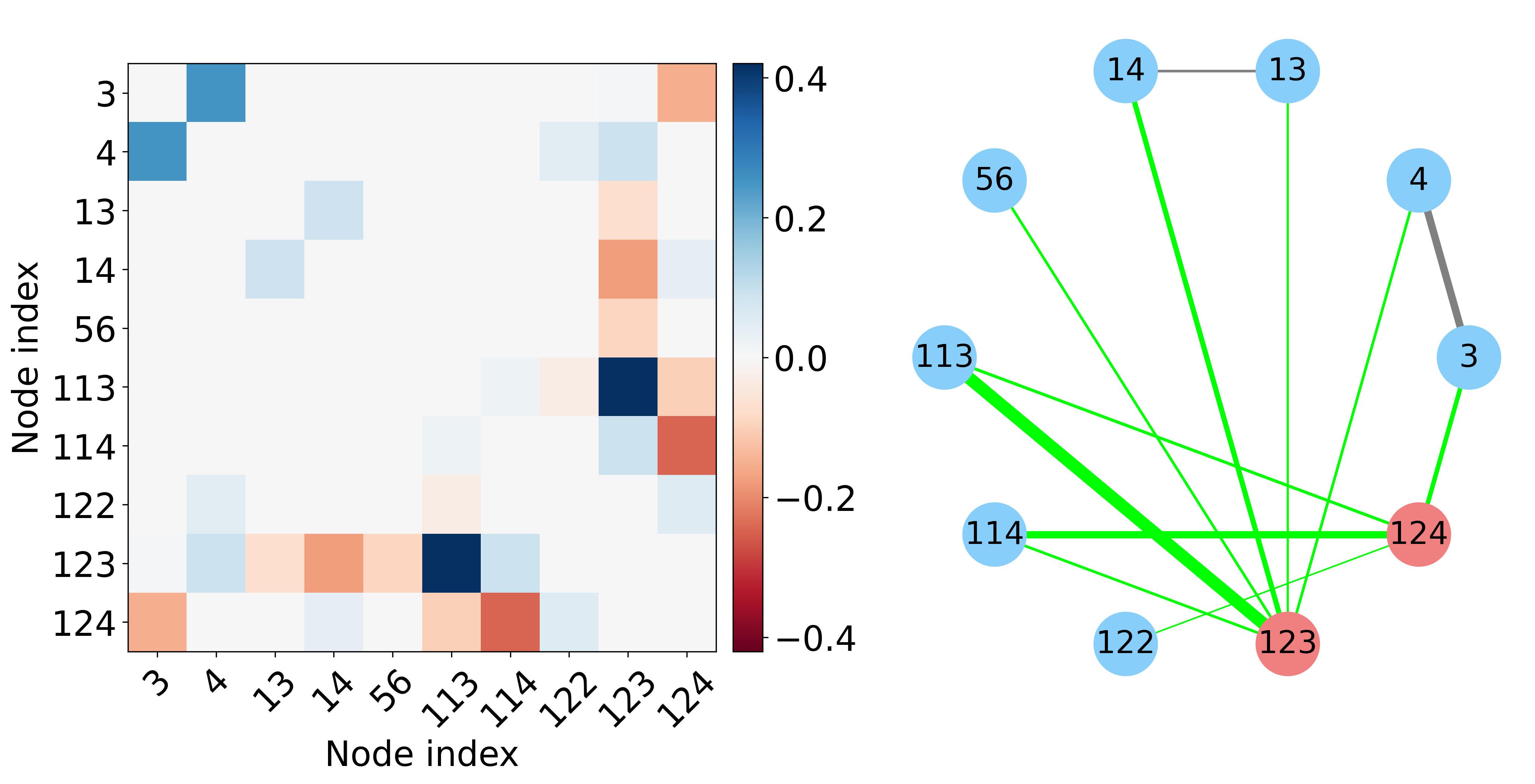}}
	\end{subfigure}
    \begin{subfigure}[Without inter-link but $\lambda=360$.]{\includegraphics[width=\columnwidth]{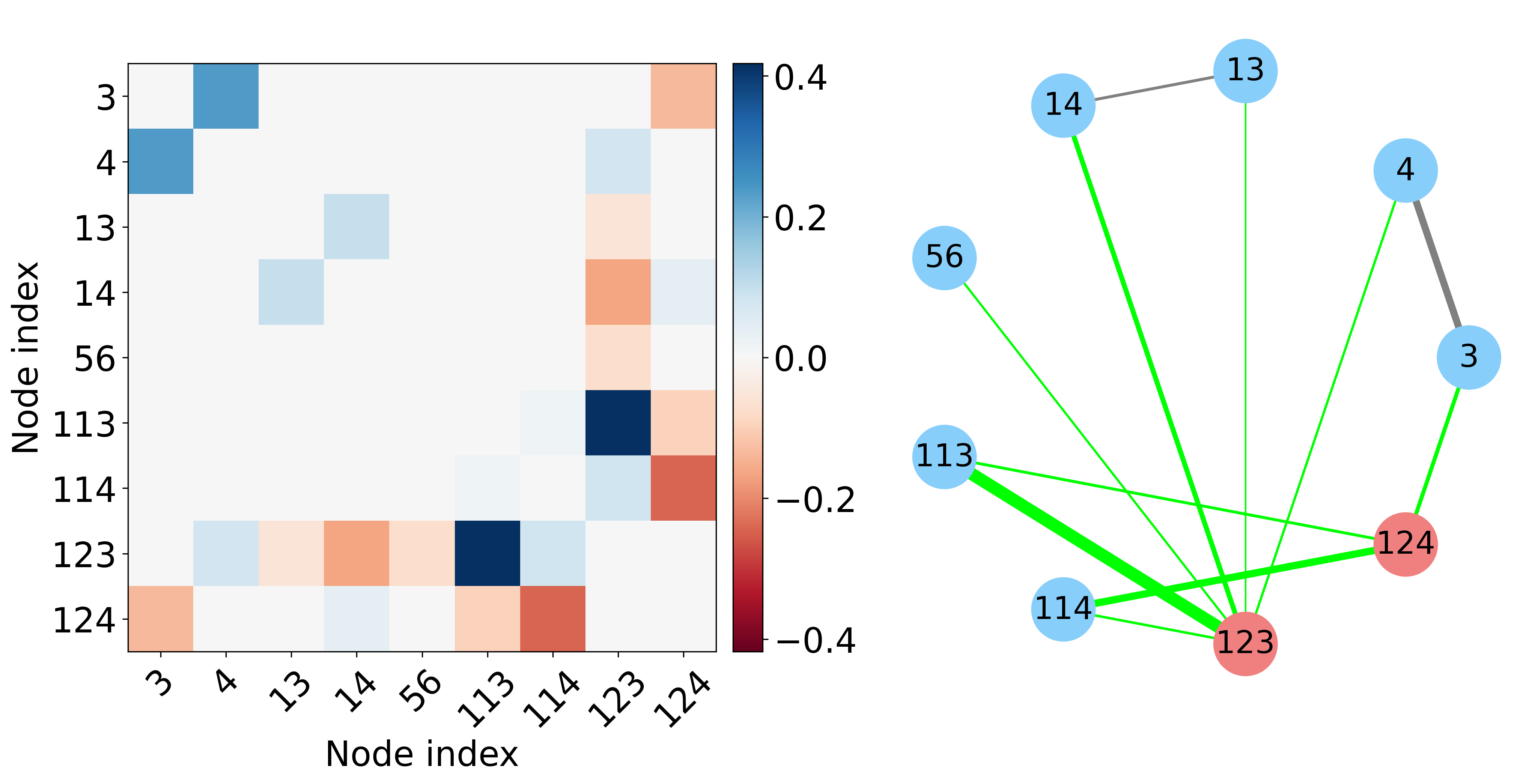}}
	\end{subfigure}
    \begin{subfigure}[Without inter-link but $\lambda=472.59$.]{\includegraphics[width=\columnwidth]{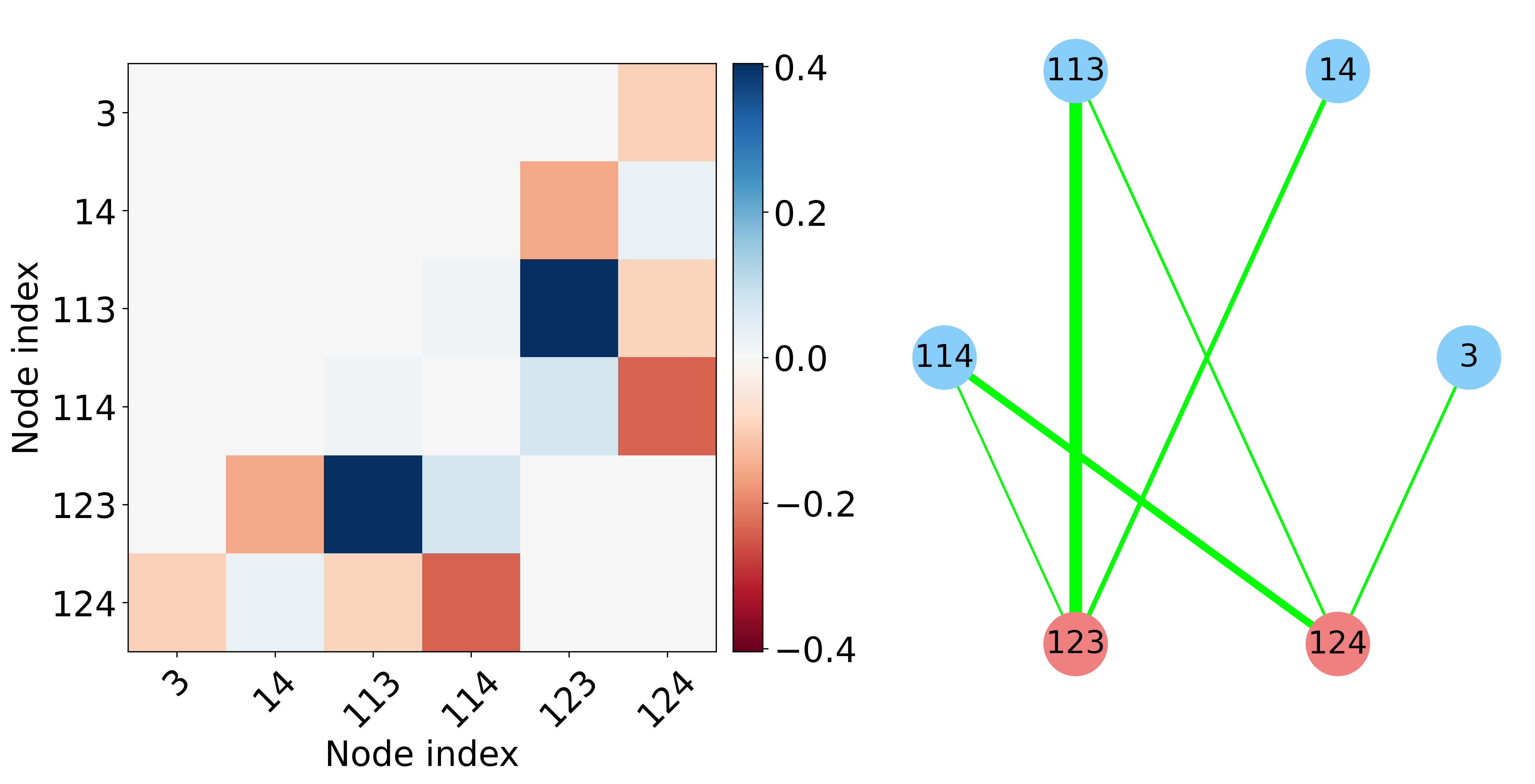}}
	\end{subfigure}
	\caption{Graph representation and associated adjacency matrix of edges involving the output variables (valence and arousal). Connectivity between output variables is not allowed. (a) Using the elbow value $\lambda=159.78$; (b) $\lambda=300$; (c) $\lambda=360$; (d) Using the largest value of $\lambda=472.59$ within our interval.}
    \label{f:graphs_without_interlink}
\end{figure*}


The graphs corresponding to the elbows of Scenarios3 and4 are depicted in Figures~\ref{FigSc3}--\ref{FigSc4}. Nodes 123 and 124 represent the two outputs, arousal and valence, respectively.
We display these graphs because they correspond to the largest elbow value ($\lambda = 159.78$) and allow for a clearer visualization of the resulting topology.
The remaining graphs in Figures~\ref{f:graphs_with_interlink}--\ref{f:graphs_without_interlink} correspond to larger values of $\lambda \in \{300,360,472.59\}$. The last value, $\lambda = 472.59$, corresponds to the extreme point of the intervals identified by the G-UAED criterion.
{We show these additional graphs to illustrate the evolution of the topology within the intervals obtained by G-UAED.}
{Figures~\ref{f:graphs_with_interlink} are devoted to the case with a link between the outputs, whereas Figures~\ref{f:graphs_without_interlink} correspond to the case without a link between the outputs.}
The information provided by the graphs is summarized in Tables~\ref{TableResultsLINK1} and~\ref{TableResultsLINK2}.
{The former considers the case with a possible inter-link between the outputs, whereas the latter reports the results without allowing this connection.
The following key points arise from a careful analysis:}

\begin{itemize}
\item The most relevant features appear to be 113 (``loudness mean'') and 114 (``loudness standard deviation''), which is consistent with previous findings in the literature. Feature~113 is typically associated with arousal, whereas feature~114 is more strongly related to the valence. Interestingly, in several graphs in Figures~\ref{f:graphs_with_interlink}--\ref{f:graphs_without_interlink}, these two features are also connected to each other.

\item When the inter-link between outputs is allowed, fewer variables are required to explain the outputs. {This suggests a connection between arousal and valence}, which is consistent with theoretical findings in psychology.

\item Valence systematically requires fewer variables than arousal. This observation contrasts with that of several previous studies reported in the literature.

\item In Table~\ref{TableResultsLINK1}, when the inter-link between outputs is allowed, valence consistently has three connections (arousal, 114, and~3), regardless of the value of $\lambda$.

\item {When valence is connected to arousal, it requires only two additional variables (114 and~3), and this configuration remains stable across variations of $\lambda$. When the connection with arousal is removed, valence requires three additional variables (six in total). Comparing Tables~\ref{TableResultsLINK1} and~\ref{TableResultsLINK2}, it appears that valence replaces the node arousal with node~113. This is reasonable because feature113 is strongly related to arousal, and features113 and114 appear to be closely connected. This suggests that arousal provides more information about valence than vice versa.}

\item Similarly, when the connection with valence is removed, arousal replaces {the node} valence with node~114, which is consistent with the previous observations.

\item From Table~\ref{TableResultsLINK1}, it can be observed that as $\lambda$ increases, node~6 disappears first, followed by node~13, and finally nodes~4 and~56.

\item From Table~\ref{TableResultsLINK2}, as $\lambda$ increases, node~6 disappears again together with node~20, followed by node~122, and finally nodes~4,~13, and~56.

\end{itemize}

\begin{table*}[!t]
\begin{center}
\caption{Connections of the two outputs (with possible inter-link) for different values of $\lambda$. {Other graph connections not involving the outputs are also reported. The connections were ordered by decreasing coefficient magnitude, providing a ranking of importance. Hence, the first column contains the indices of the most important features in each case.} {\bf A: Arousal, V: Valence.}} \label{TableResultsLINK1}
{
\begin{tabular}{|c|| c c c c c c c c||c||c|} 
 \hline
\multirow{ 2}{*}{ {\bf Output}} 
 & \multicolumn{8}{c||}{{\bf Connections}} & \multirow{ 2}{*}{$\lambda$}  & \multirow{ 2}{*}{ {\bf Other connections}} \\ 
  \cline{2-9}
    &  1 &  2  & 3  & 4 & 5 & 6 & 7 & 8  &  & \\ 
 \hline
  \hline
{\bf Arousal (A)}  & 113  & V   & 56& 14 & 13 & 6 & 4 & & \multirow{ 2}{*}{159.78} &  \multirow{ 2}{*}{14-13, 14-6, 6-3, 114-113 and 4-3} \\ 
{\bf Valence (V)}   & A & 114 & 3 & &  & & & & & \\ 
  \hline
  \hline
{\bf Arousal (A)}  & 113  & V   & 56& 14 & 13 & 4 &  & & \multirow{ 2}{*}{300} &  \multirow{ 2}{*}{14-13, 114-113 and 4-3} \\ 
{\bf Valence (V)}   & A & 114 & 3 & &  & & & & & \\ 
  \hline
  \hline
{\bf Arousal (A)}  & 113  & V   & 56& 14 & 4 &  & & &\multirow{ 2}{*}{360} &  \multirow{ 2}{*}{114-113 and 4-3} \\ 
{\bf Valence (V)}   & A & 114 & 3 & &  & & & & & \\ 
 \hline
  \hline
{\bf Arousal (A)}  & 113  & V   &  14 & & &  & & &\multirow{ 2}{*}{472.59} &  \multirow{ 2}{*}{114-113} \\ 
{\bf Valence (V)}   & A & 114 & 3 & &  & & & & & \\ 
 \hline
\end{tabular}
}
\end{center}
\end{table*}

\begin{table*}[!t]
\begin{center}
\caption{Connections of the two outputs (without possible inter-link) for different values of $\lambda$. {Other graph connections not involving the outputs are also reported. The connections were ordered by decreasing coefficient magnitude, providing a ranking of importance. Thus, the first column contains the indices of the most important features in each case.} {\bf A: Arousal, V: Valence.}} \label{TableResultsLINK2}
{
\begin{tabular}{|c|| c c c c c c c c||c||c|} 
 \hline
\multirow{ 2}{*}{ {\bf Output}} 
 & \multicolumn{8}{c||}{{\bf Connections}} & \multirow{ 2}{*}{$\lambda$}  & \multirow{ 2}{*}{ {\bf Other connections}} \\ 
  \cline{2-9}
    &  1 &  2  & 3  & 4 & 5 & 6 & 7 & 8  &  & \\ 
 \hline
  \hline
{\bf Arousal (A)}  & 113  & 14   & 13 & 56 & 114 & 4 & 6 & & \multirow{ 2}{*}{159.78} &  \multirow{ 2}{*}{20-14, 14-13, 6-3, 114-113, 122-4 and 4-3} \\ 
{\bf Valence (V)}   & 114 & 3& 113 & 122 & 20 & 4 & & & & \\ 
  \hline
  \hline
{\bf Arousal (A)}  & 113  & 14   & 56& 13 & 114 & 4 &  & & \multirow{ 2}{*}{300} &  \multirow{ 2}{*}{14-13 and 4-3} \\ 
{\bf Valence (V)}   & 114 & 3 & 113   & 122 &  & & & & & \\ 
  \hline
  \hline
{\bf Arousal (A)}  & 113  & 14   & 56& 13 & 4 & 114 & & &\multirow{ 2}{*}{360} &  \multirow{ 2}{*}{14-13 and 4-3} \\ 
{\bf Valence (V)}   & 114 & 113 & 3 & &  & & & & & \\ 
 \hline
  \hline
{\bf Arousal (A)}  & 113  & 14   &  114 & & &  & & &\multirow{ 2}{*}{472.59} &  \multirow{ 2}{*}{-----} \\ 
{\bf Valence (V)}   & 114 & 113 & 3 & &  & & & & & \\ 
 \hline
\end{tabular}
}
\end{center}
\end{table*}

{These observations can be further interpreted by comparing them with previous results reported in the literature.}
{In} \cite{OurPaperSound}, a linear regression method was also applied to this dataset jointly with wrapper methods and a Gibbs analysis for feature selection, finding that (see \cite[Section 6]{OurPaperSound}) the most relevant variables in common to the two outputs were 114, 113, 14, 115 (``energy mean''), and 3. The frequency-domain features 14 (``spread mean'') and 3 (``decrease slope mean'') also appear in our graphs and Tables~\ref{TableResultsLINK1}--\ref{TableResultsLINK2}. Thus, with the exception of {feature} 115, {the two studies are in agreement in this respect}. Note that variable 4 (``maximum fluctuation'') is quite connected to 3, according to our graphs.
Furthermore, in \cite{OurPaperSound}, the authors suggest a linear model with $7$ variables, 113, 114, 14, 115, 4, 8 (``roll-off mean''), and 56 (``pitch standard deviation''), for arousal. If we compare {this result with} Table~\ref{TableResultsLINK2} {for the value of} $\lambda$ corresponding to the elbow, i.e., $159.78$, we {also obtain} $7$ variables, which include $113$, $114$, $14$, $4$, {and} $56$, but {with} $6$ and $13$ instead of $115$ and $8$. However, there is {agreement on} variables: $113$, $114$, $14$, $4$, and $56$.
On the other hand, regarding valence, \cite{OurPaperSound} {suggests} a model with $16$ variables, whereas, {looking at} Table~\ref{TableResultsLINK2} with $\lambda=159.78$ {and without the connection to arousal}, {our results suggest} at most $6$ variables for valence: 114, 3, 113, 122, 20, and 4. However, all these features, {except for} variable 122, are contained in the linear model with $16$ variables in \cite{OurPaperSound}.
 
{The results with larger values of $\lambda$ are in line with the results obtained in the literature using nonlinear regression methods which, in general, suggest using a smaller number of features in the model}. For instance, the analysis with regression trees in \cite{DTR_Access_24} suggests the use of $2$ variables for arousal, {namely} 113 (``Loudness mean'') and 4 (``Fluctuation max''), which also appear in our graphs and Tables~\ref{TableResultsLINK1}--\ref{TableResultsLINK2}.
Regarding valence, the study in \cite{DTR_Access_24} suggests a nonlinear regression model based on $4$ variables: 114 (``Loudness standard deviation''), 113 (``Loudness mean''), 3 (``Decrease Slope mean''), and 115 (``energy mean''). With the exception of the latter (feature 115), {
results in Table~\ref{TableResultsLINK2} for $\lambda=472.59$ (which is one of the extreme values obtained by G-UAED) shows a strong agreement.}

\begin{figure*}[!t]
  \centering
  \centerline{
      \subfigure{\includegraphics[width=0.45\columnwidth]{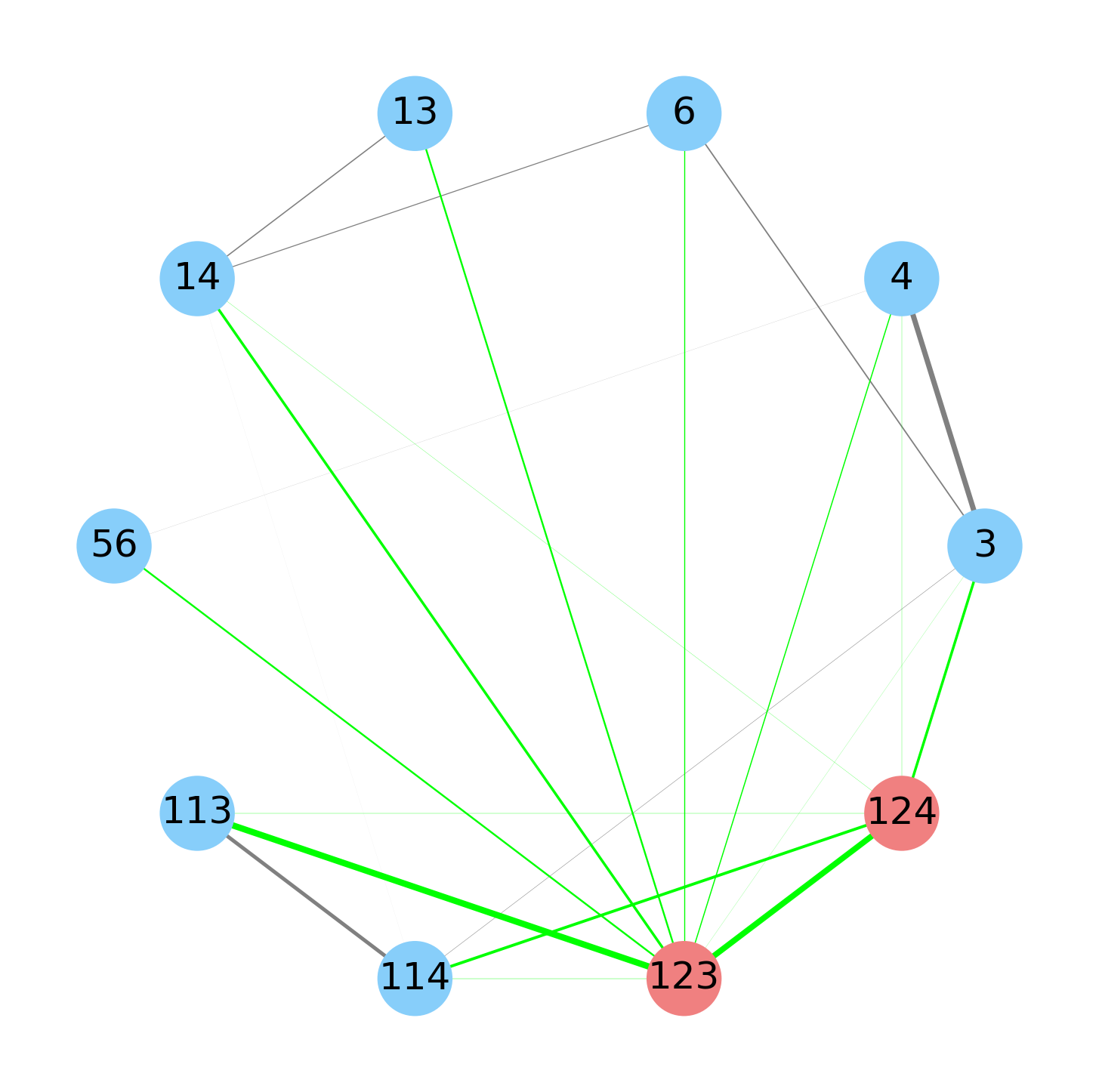}}
      \subfigure{\includegraphics[width=0.45\columnwidth]{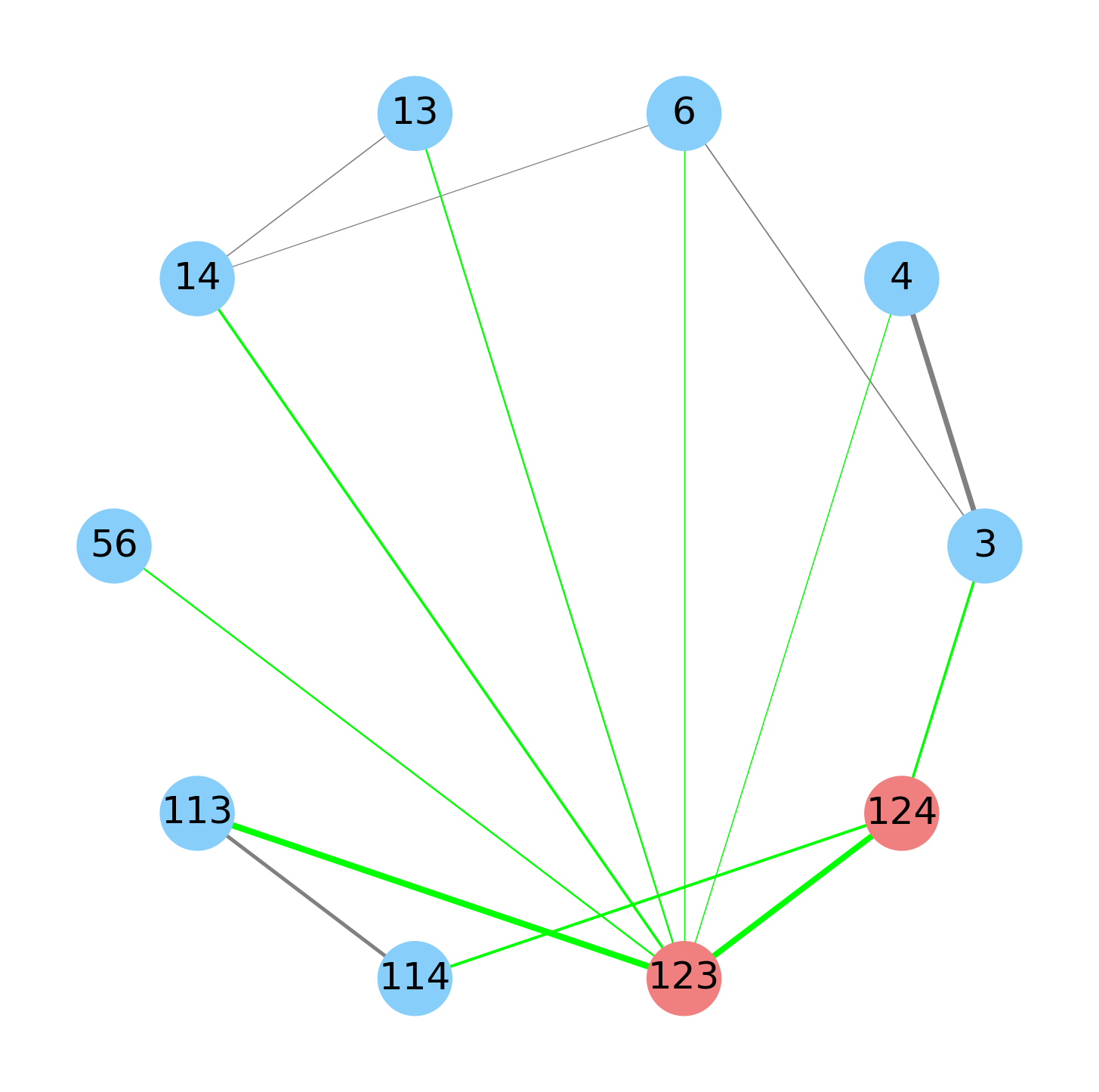}}
      \subfigure{\includegraphics[width=0.45\columnwidth]{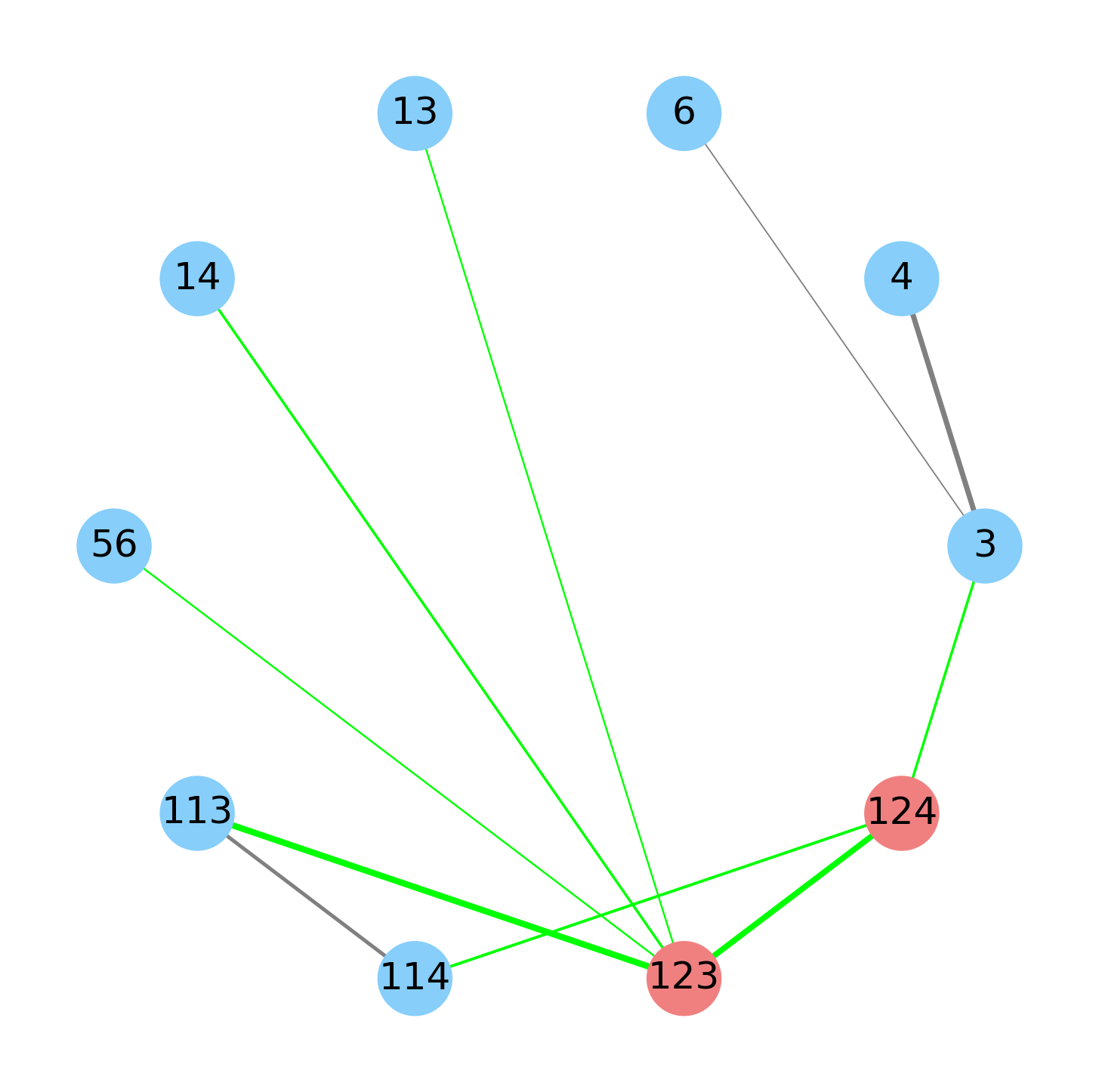}}
      \subfigure{\includegraphics[width=0.45\columnwidth]{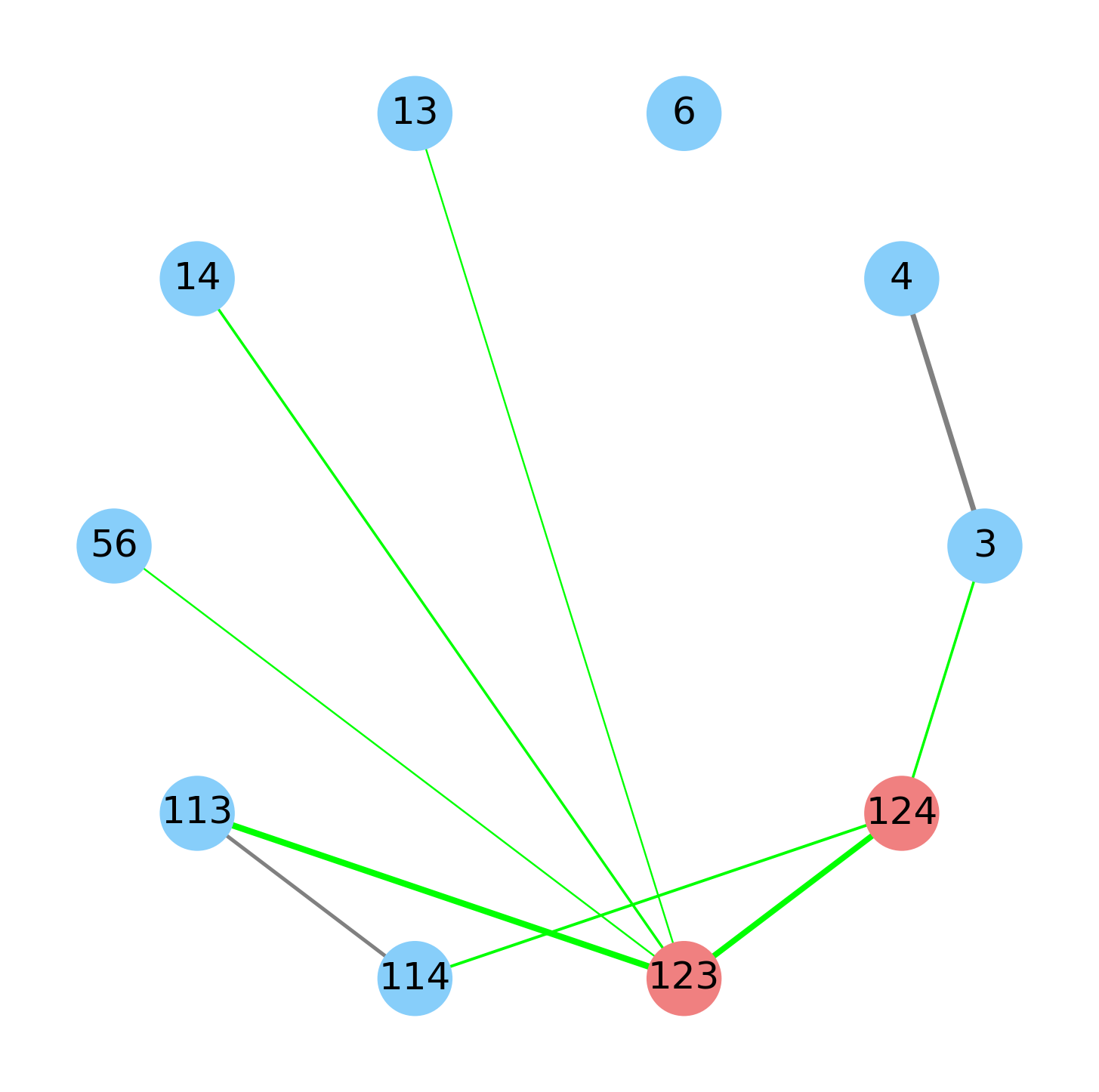}}
  }
  \setcounter{subfigure}{0} 
  \centerline{
      \subfigure[\label{th0_159} $\delta = 0$]{\includegraphics[width=0.45\columnwidth]{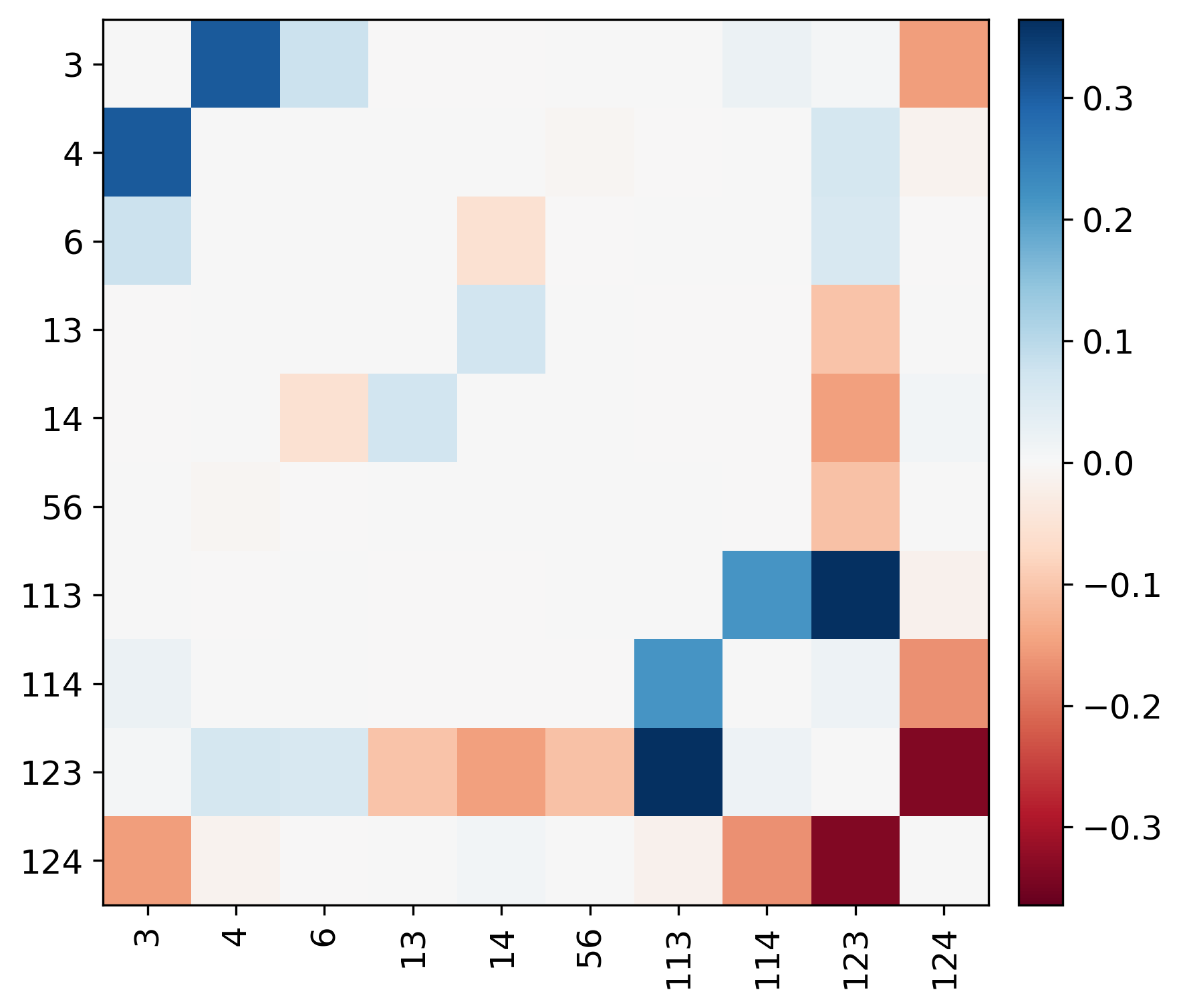}}
      \subfigure[\label{th05_159} $\delta = 0.05$]{\includegraphics[width=0.45\columnwidth]{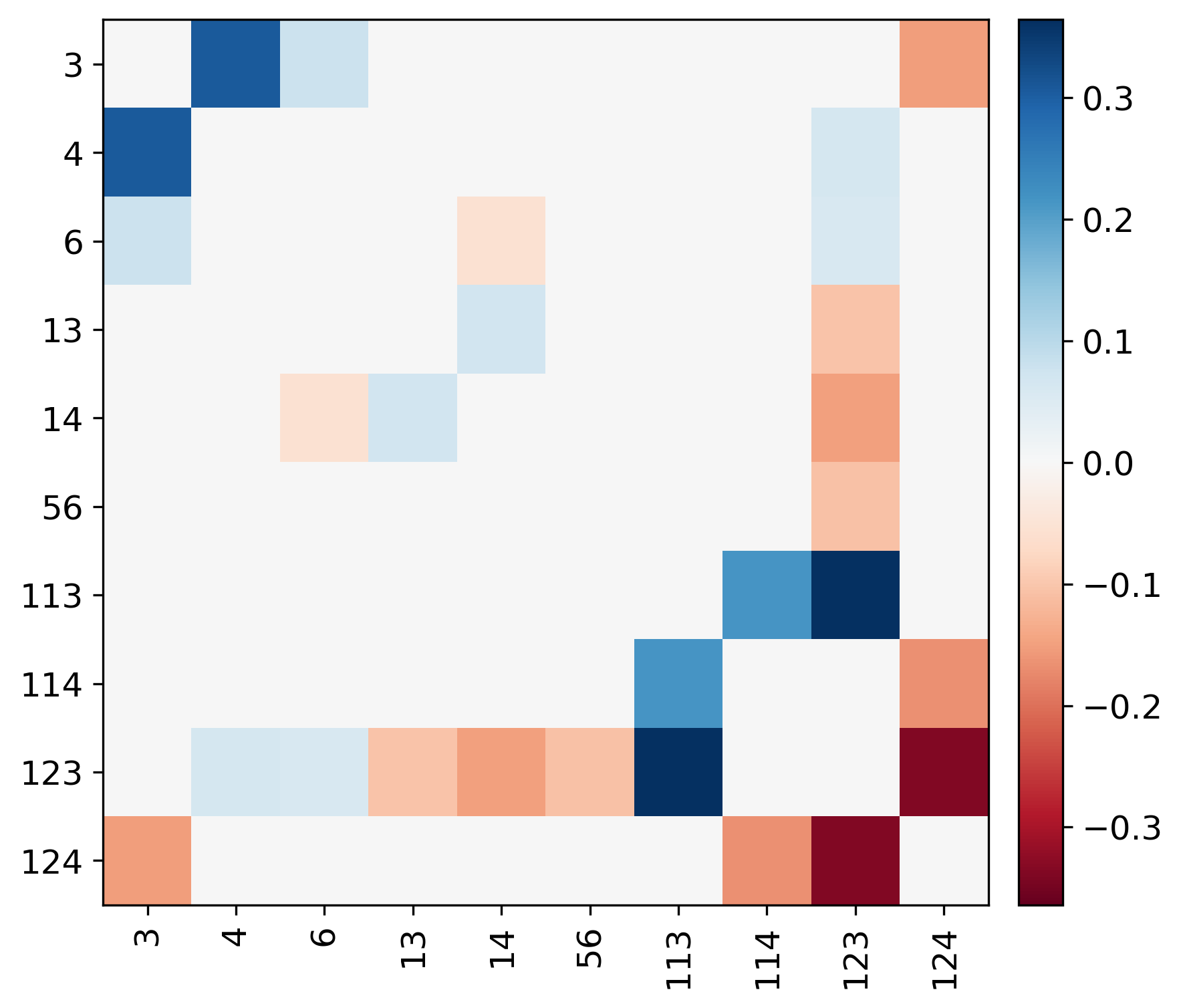}}
      \subfigure[\label{th075_159} $\delta = 0.075$]{\includegraphics[width=0.45\columnwidth]{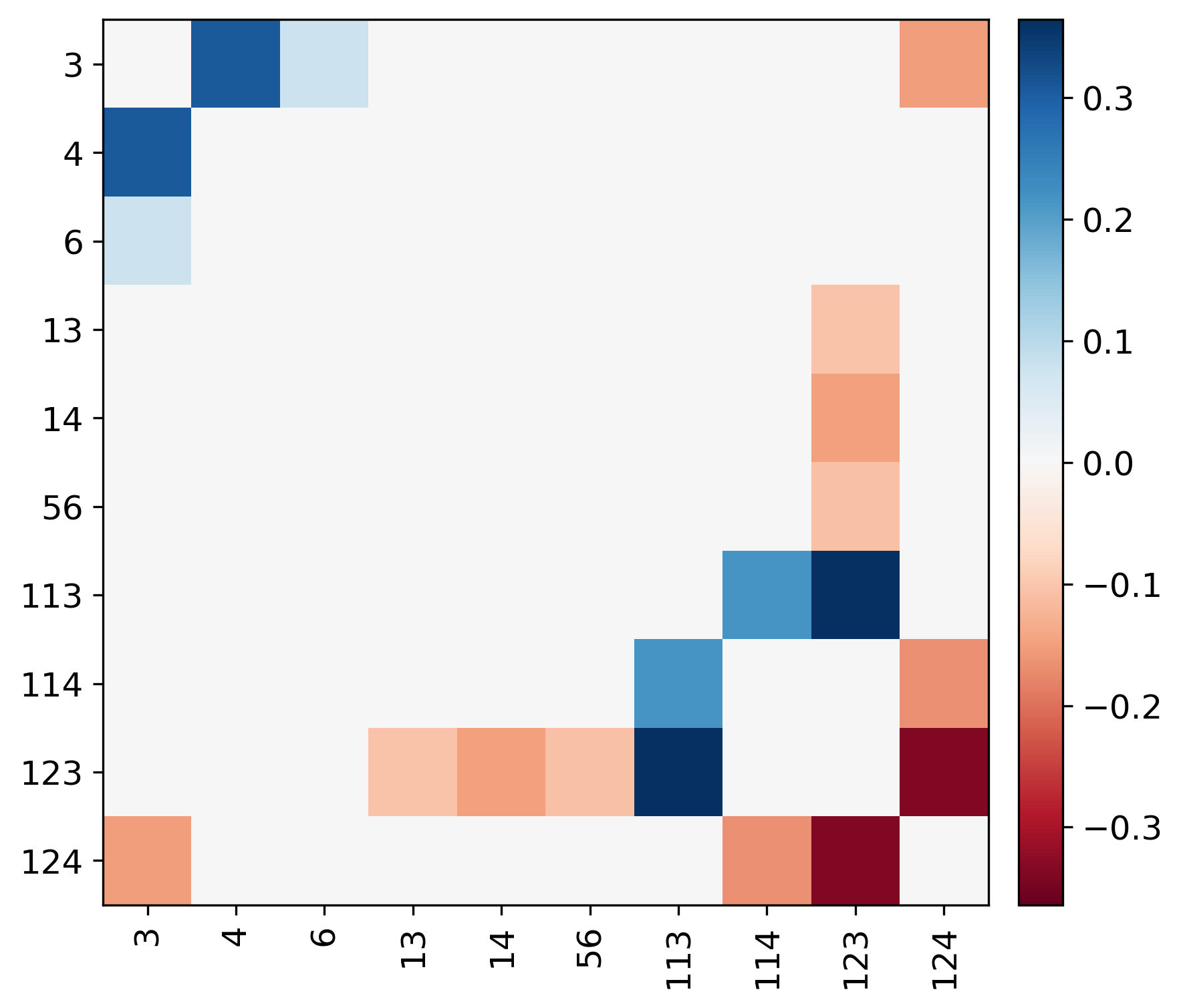}}
      \subfigure[\label{th1_159} $\delta = 0.1$]{\includegraphics[width=0.45\columnwidth]{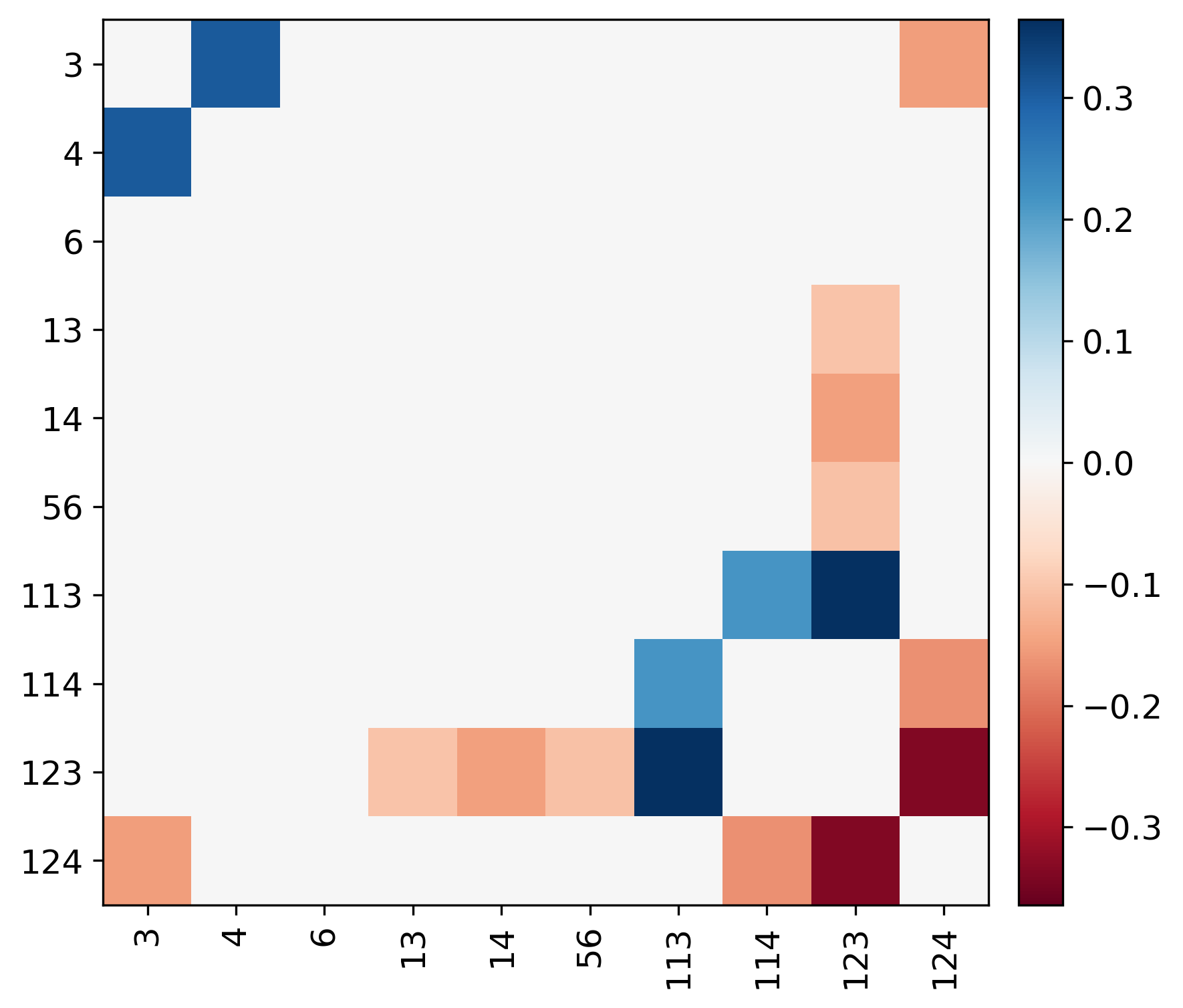}}
  }
  \centerline{
      \subfigure{\includegraphics[width=0.45\columnwidth]{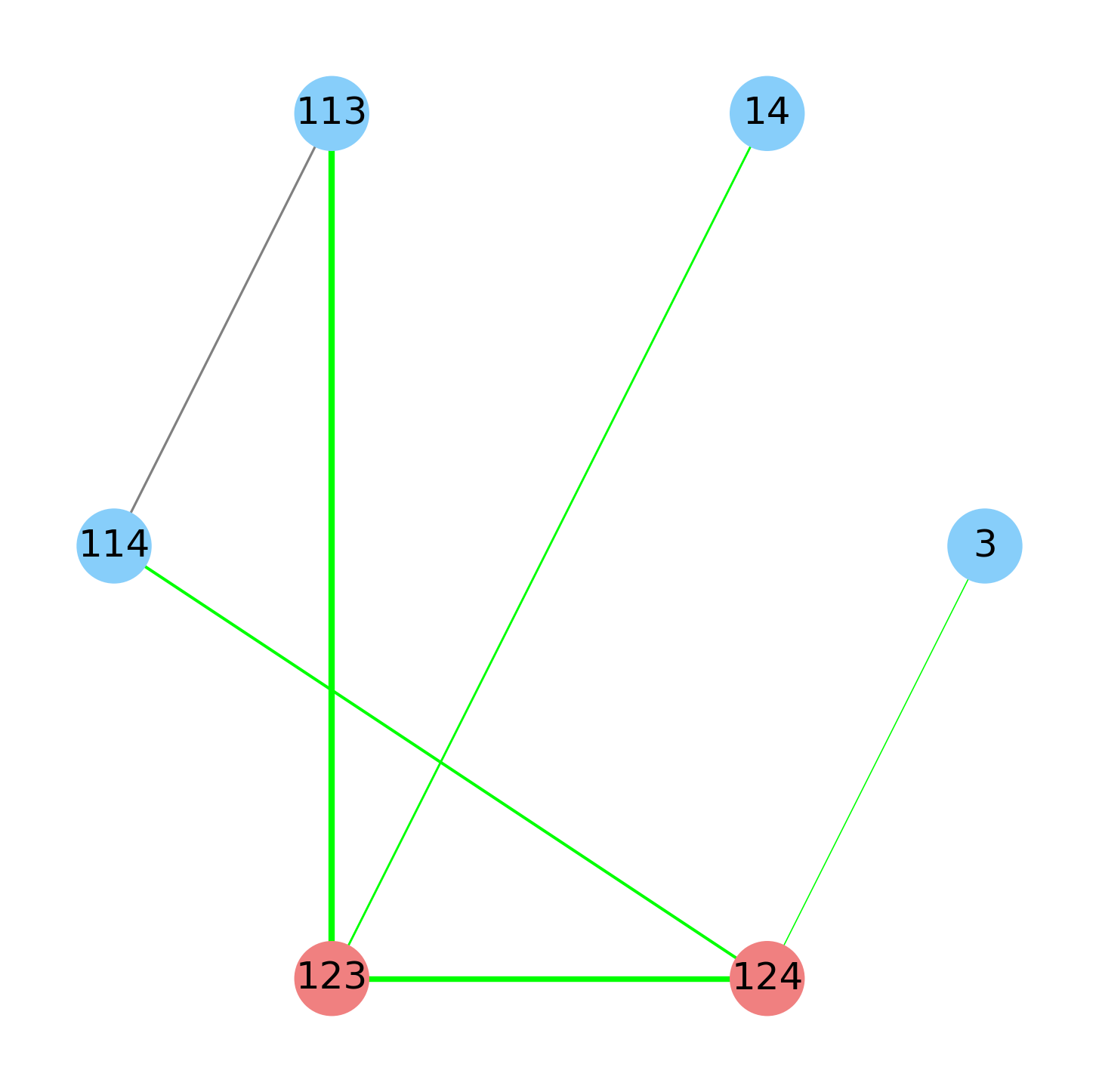}}
      \subfigure{\includegraphics[width=0.45\columnwidth]{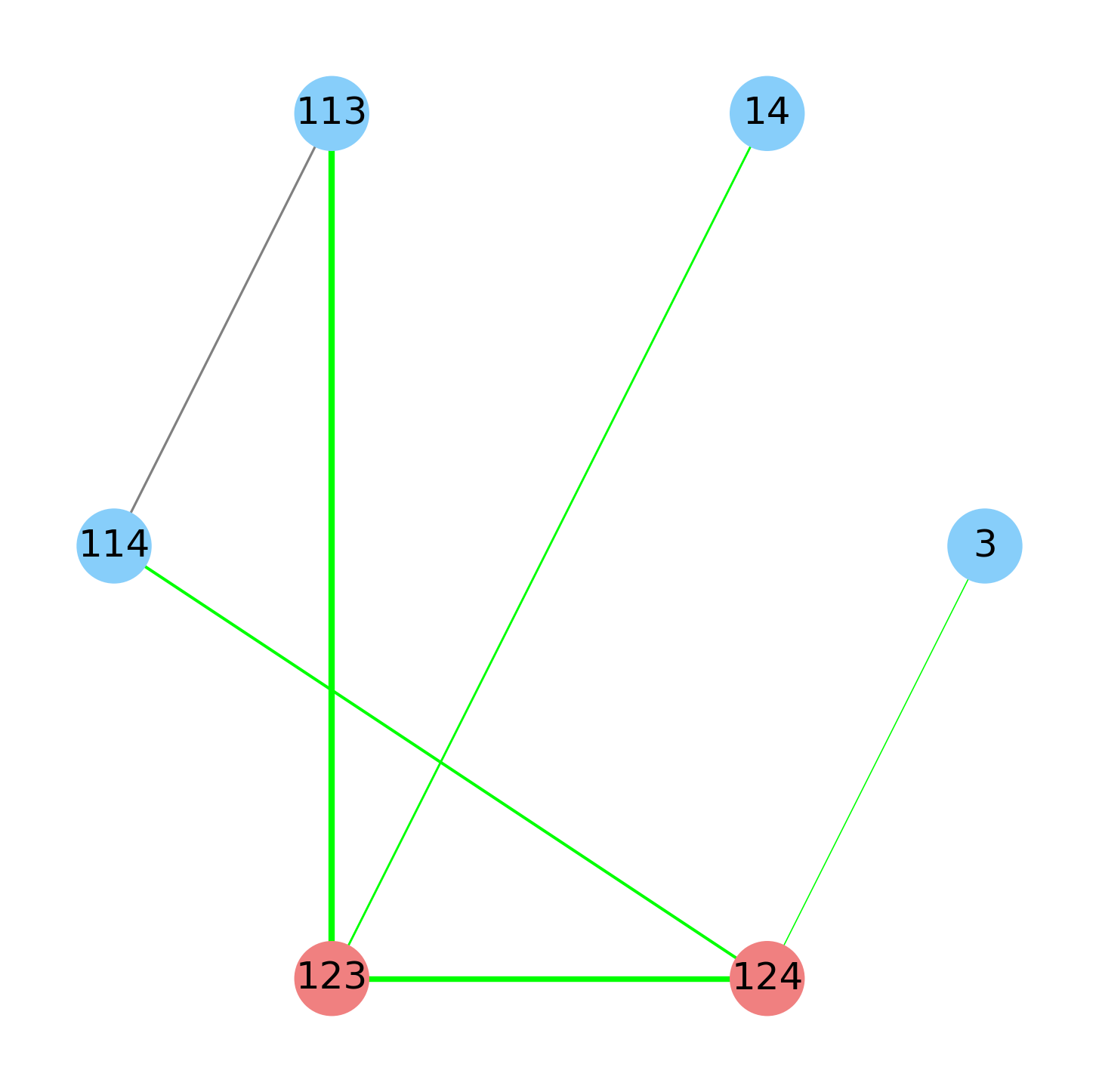}}
      \subfigure{\includegraphics[width=0.45\columnwidth]{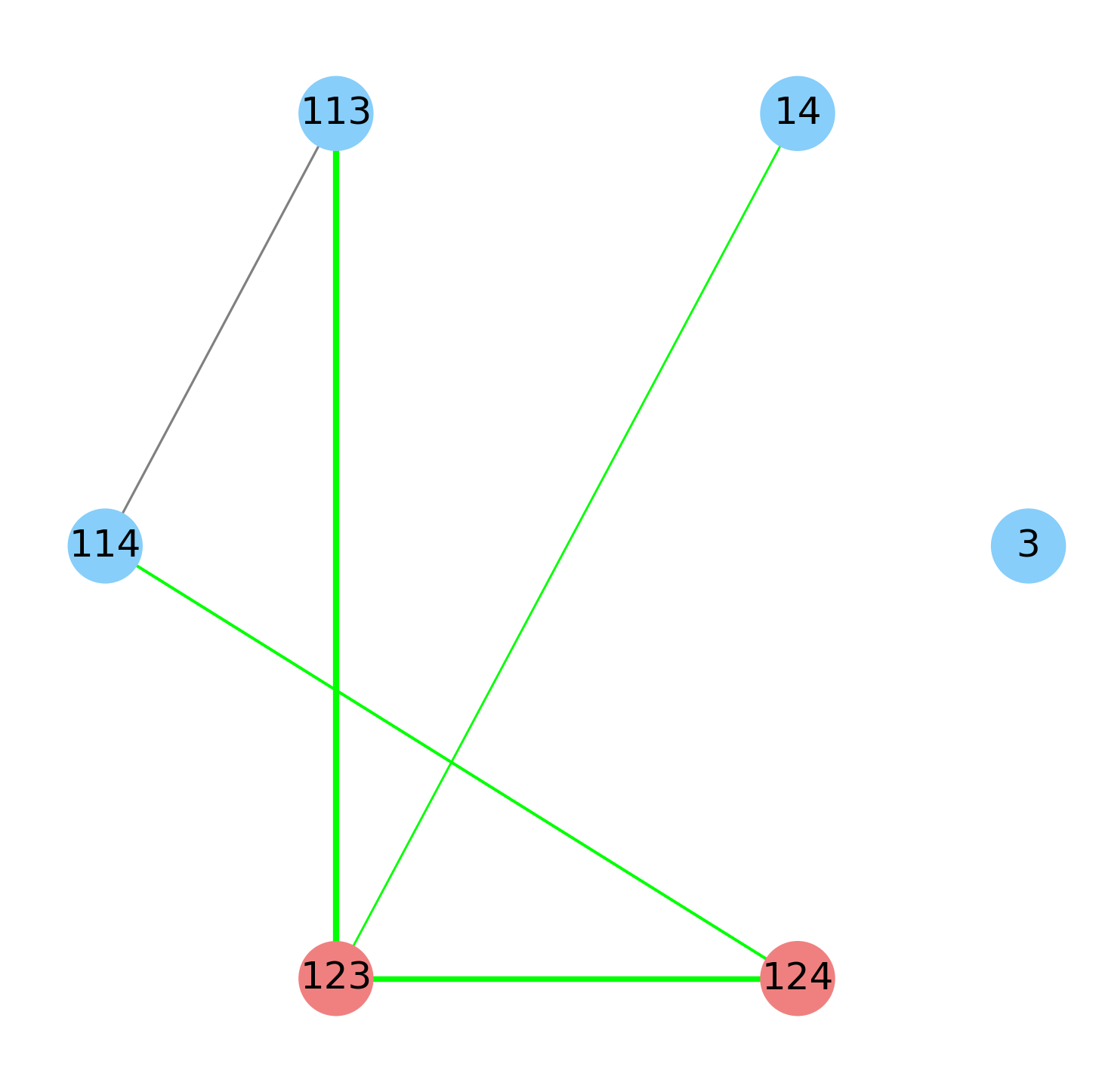}}
      \subfigure{\includegraphics[width=0.45\columnwidth]{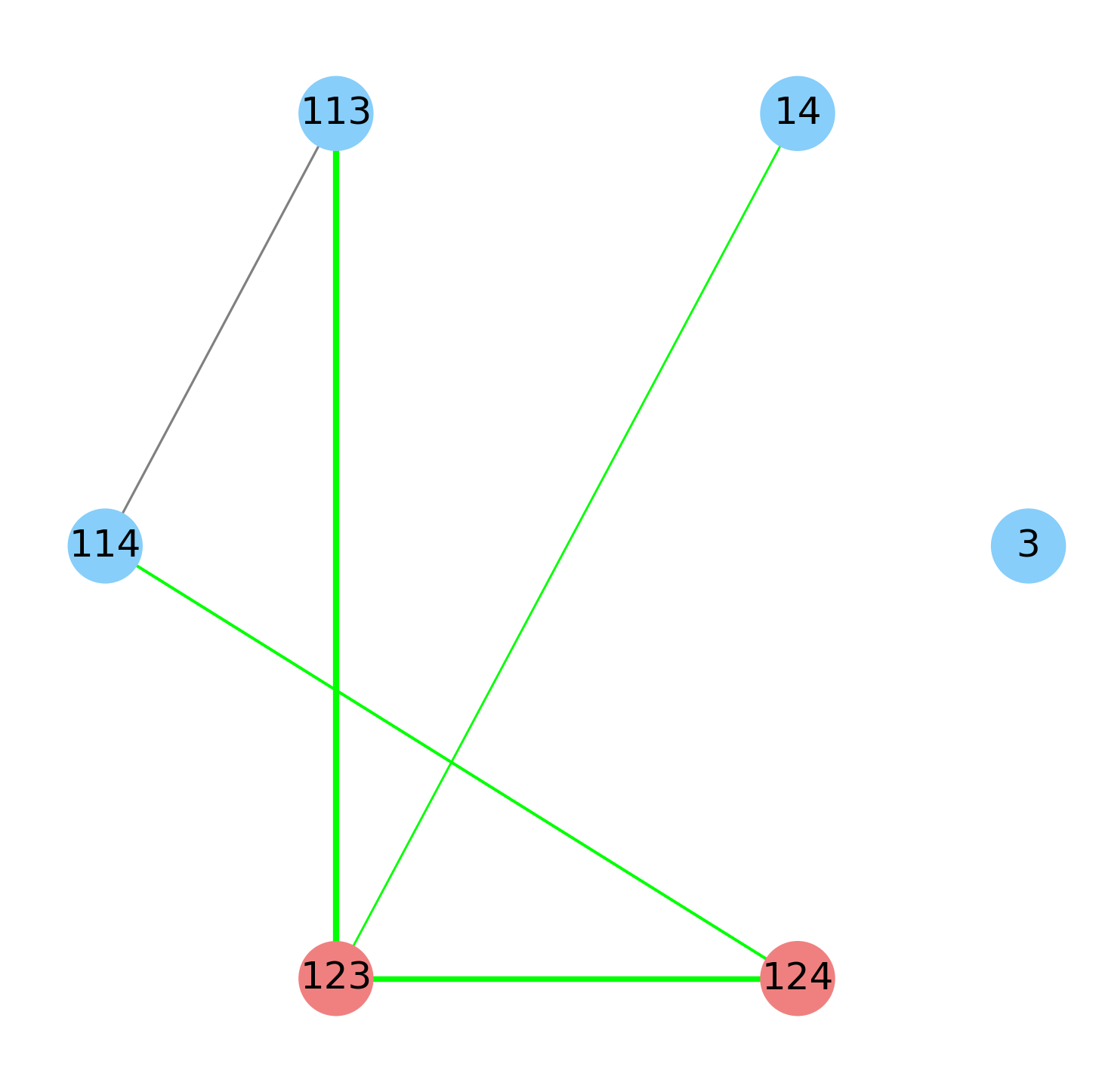}}
  }
  \setcounter{subfigure}{4} 
  \centerline{
      \subfigure[\label{th0_472} $\delta = 0$]{\includegraphics[width=0.45\columnwidth]{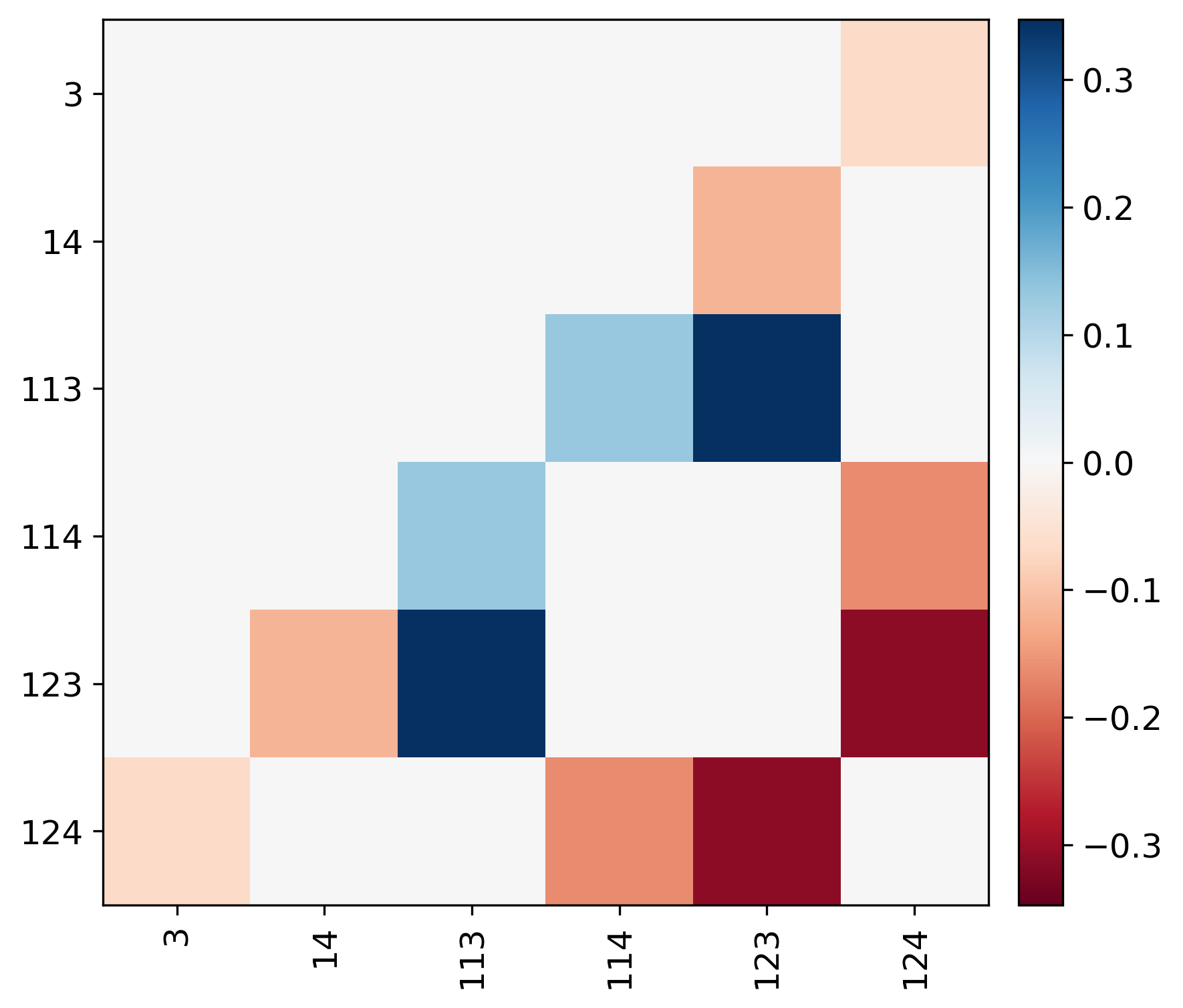}}
      \subfigure[\label{th05_472} $\delta = 0.05$]{\includegraphics[width=0.45\columnwidth]{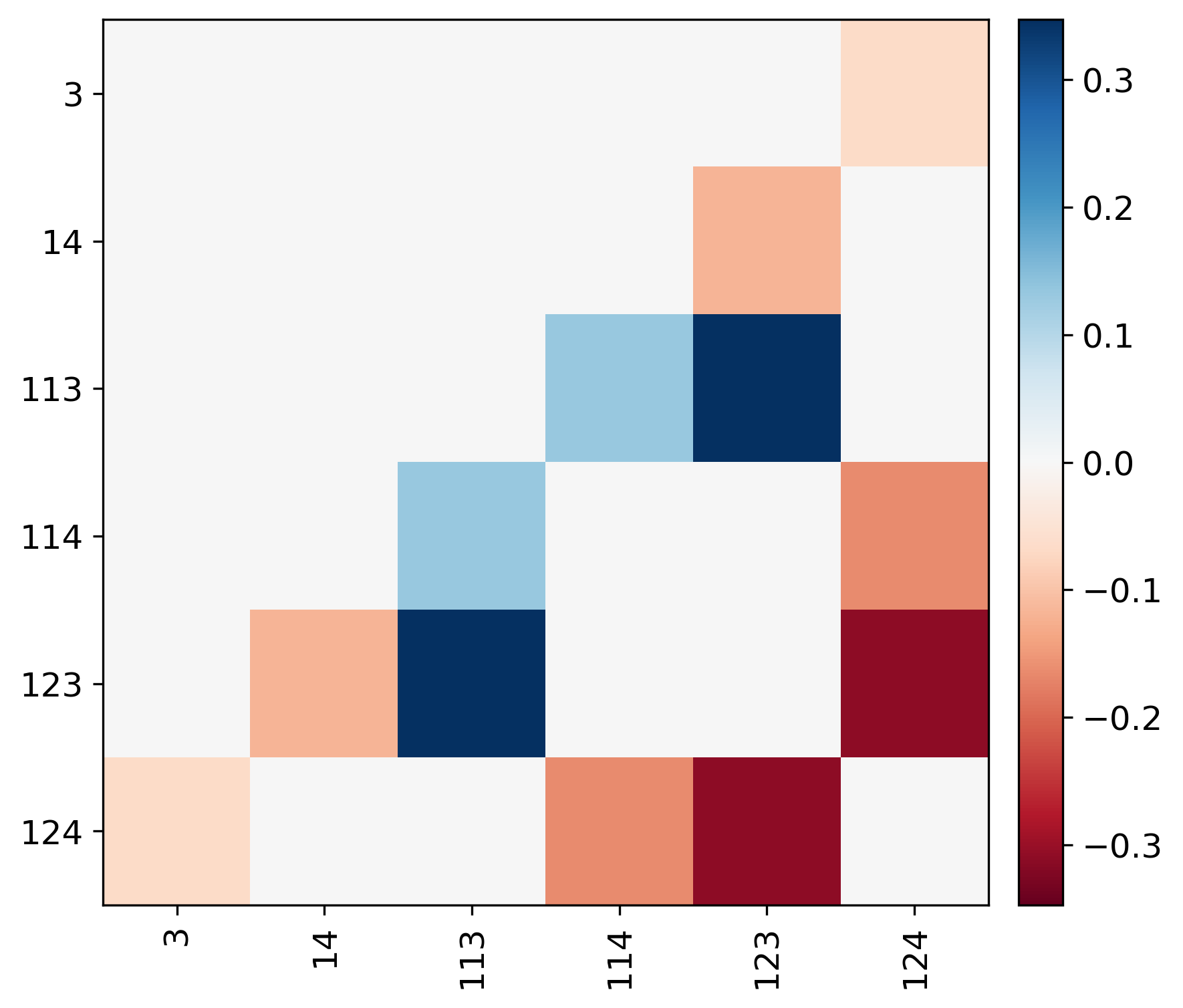}}
      \subfigure[\label{th075_472} $\delta = 0.075$]{\includegraphics[width=0.45\columnwidth]{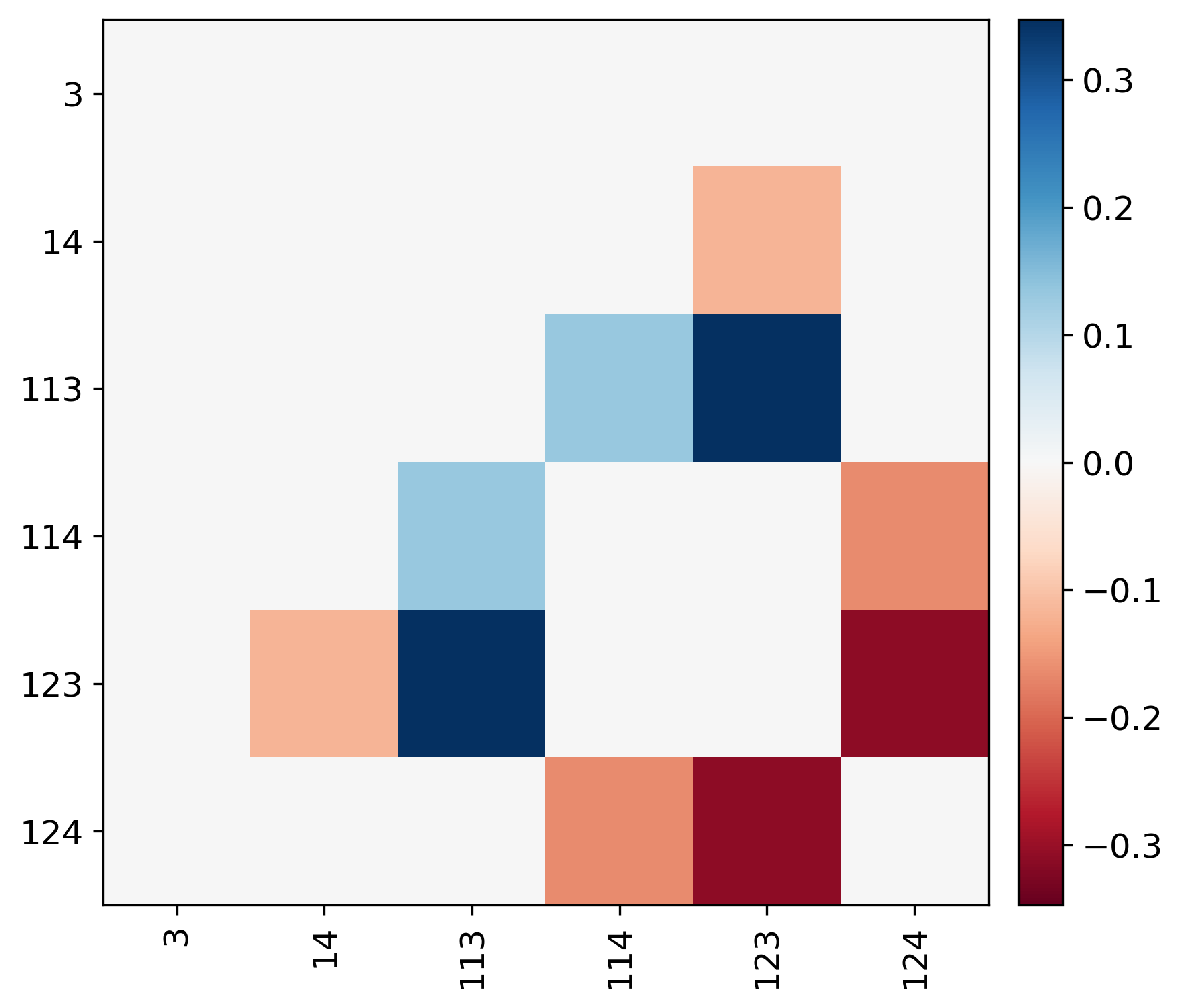}}
      \subfigure[\label{th1_472} $\delta = 0.1$]{\includegraphics[width=0.45\columnwidth]{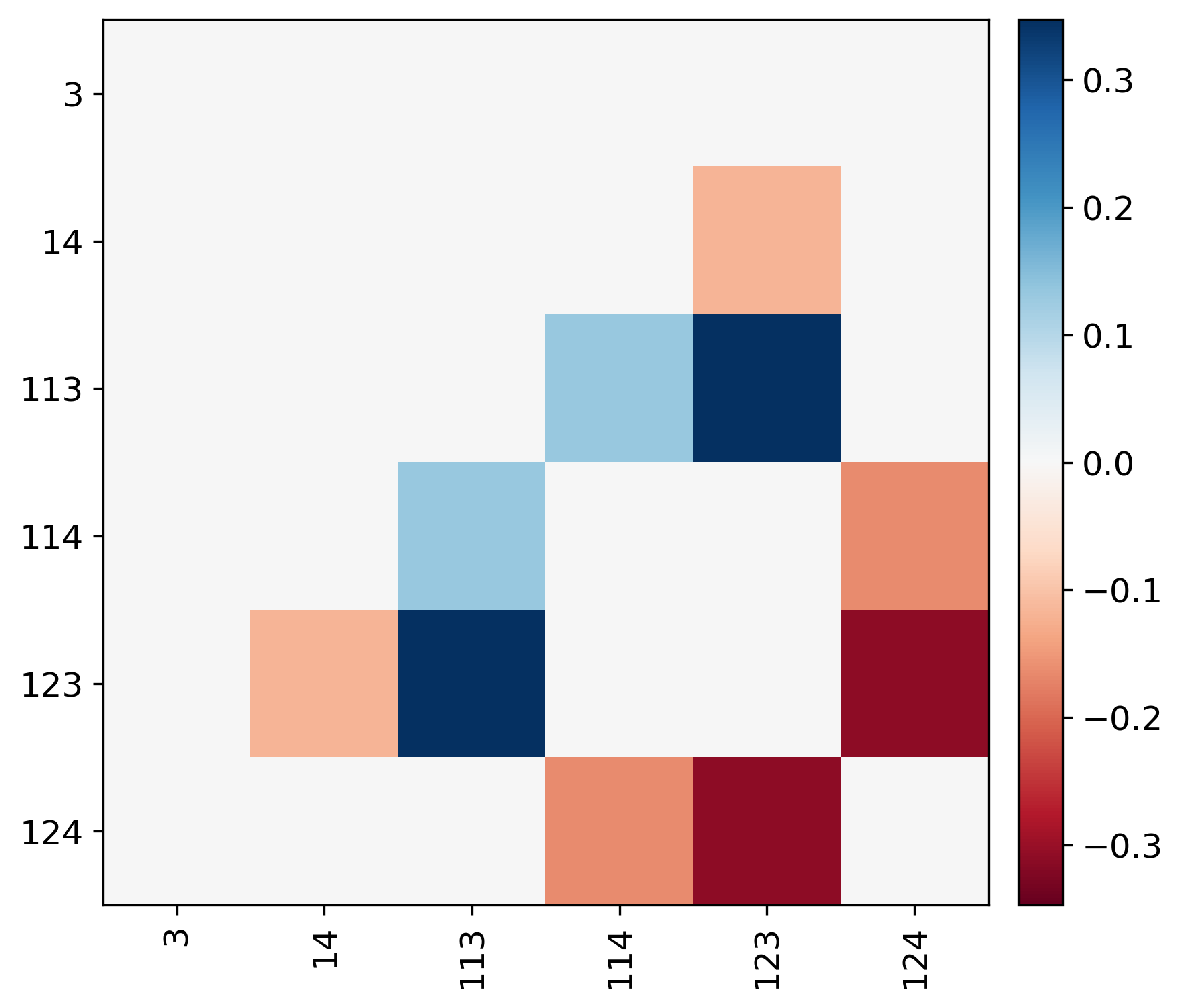}}
  }
  \caption{{Sensitivity analysis with respect to the threshold $\delta$ for $\lambda = 159.78$ [panels (a)–(d)] and $\lambda = 472.59$ [panels (e)–(h)]. For each value of $\lambda$, the top row displays the inferred graphs after thresholding and the bottom row the corresponding adjacency matrices. Results show that dominant connections are preserved across a broad range of threshold values, with visible changes only for larger $\delta$.}}
  \label{fig:thresholds}
\end{figure*}

{\vspace{2mm} \noindent
\textbf{Influence of the threshold $\delta$ (ablation study).}
To assess the robustness of the inferred topology with respect to the post-processing threshold, we conducted an ablation study examining the impact of different values of $\delta$ on the learned graph. 
We consider two representative values of the regularization parameter, $\lambda = 159.78$ and $\lambda = 472.59$, which correspond, respectively, to the elbow and the upper bound for $\lambda$ in Table~\ref{TableElbow}, and analyze the graphs obtained for different threshold values of $\delta$. The considered values of $\delta$ were selected according to the magnitude of the maximum edge weight.
Figure~\ref{fig:thresholds} illustrates the corresponding graphs (top row) and the adjacency matrices (bottom row). Figures~(a)--(d) correspond to $\lambda = 159.78$, whereas Figures~(e)--(h) correspond to $\lambda = 472.59$, again illustrating that larger values of $\lambda$ correspond to fewer connections.  
The results indicate that the estimated graphs remain stable across a broad range of threshold values because the main links are preserved for all considered values of $\delta$. Noticeable changes appear only for larger thresholds (e.g., $\delta = 0.1$, which corresponds to approximately one-third of the maximum edge weight). 
Furthermore, the results highlight the expected interaction between $\lambda$ and $\delta$: larger values of $\lambda$ naturally yield sparser graphs and therefore require smaller thresholds to obtain subgraph structures that are comparable. This can be observed, for instance, by comparing the subgraphs in Figure~\ref{fig:thresholds}(d) and Figure~\ref{fig:thresholds}(e).
}

\subsection{{Feature selection on the ARAUS database (experiment)}}

\begin{figure}[!ht]  
    \centering
    \includegraphics[width=0.8\columnwidth]{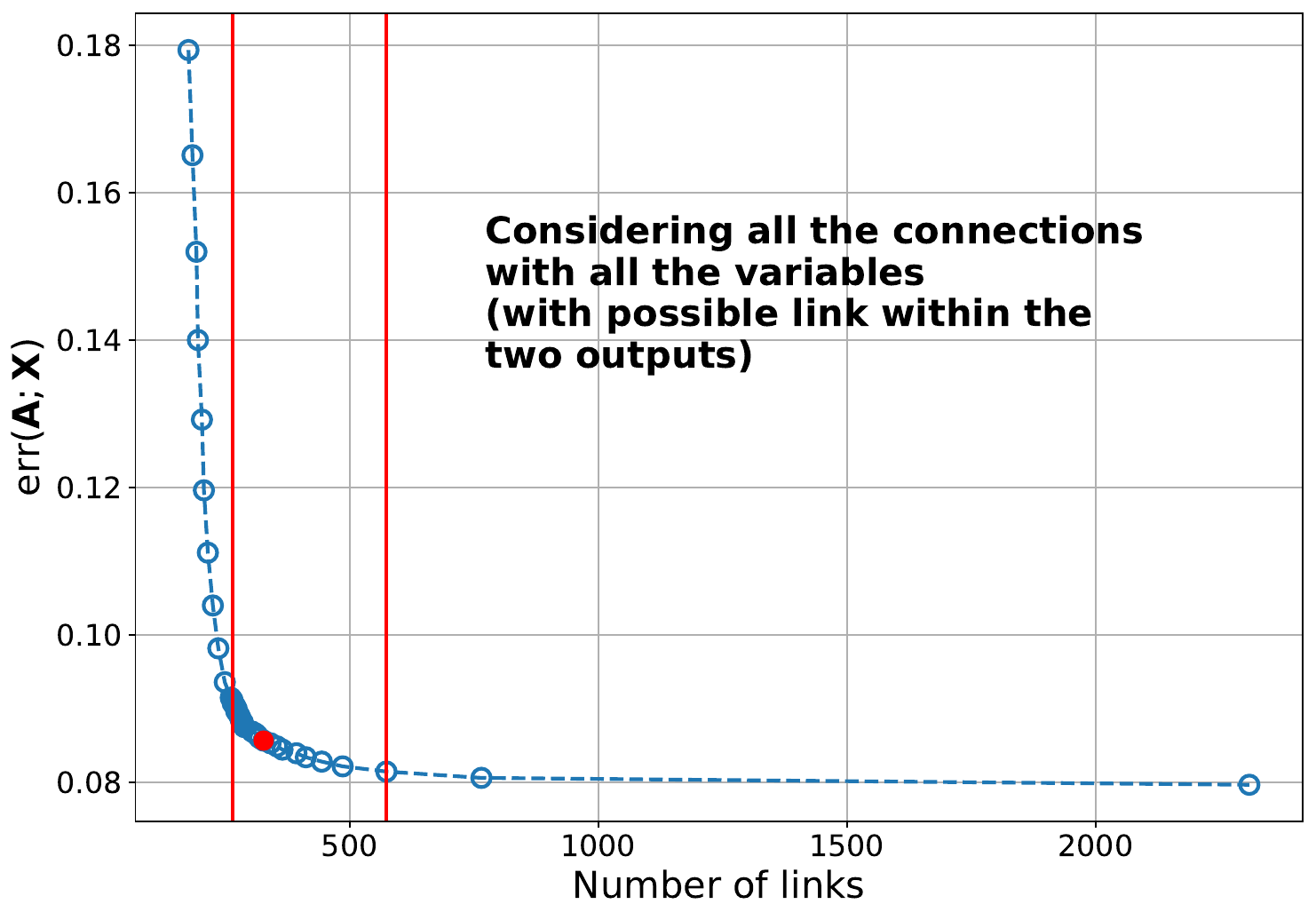}
    \caption{{Normalized error as a function of the number of links in the graph for the ARAUS dataset in Scenario~1, where the connection between outputs is allowed. The red points and red lines indicate the elbows and intervals obtained by applying G-UAED as described in Algorithm~\ref{alg:sem_guaed}.}}
   \label{fig:elbow_araus}
\end{figure}

\begin{figure*}[!t]
	\centering
	\begin{subfigure}[Scenario 1: $\lambda=0.56$.]{\includegraphics[width=\columnwidth]{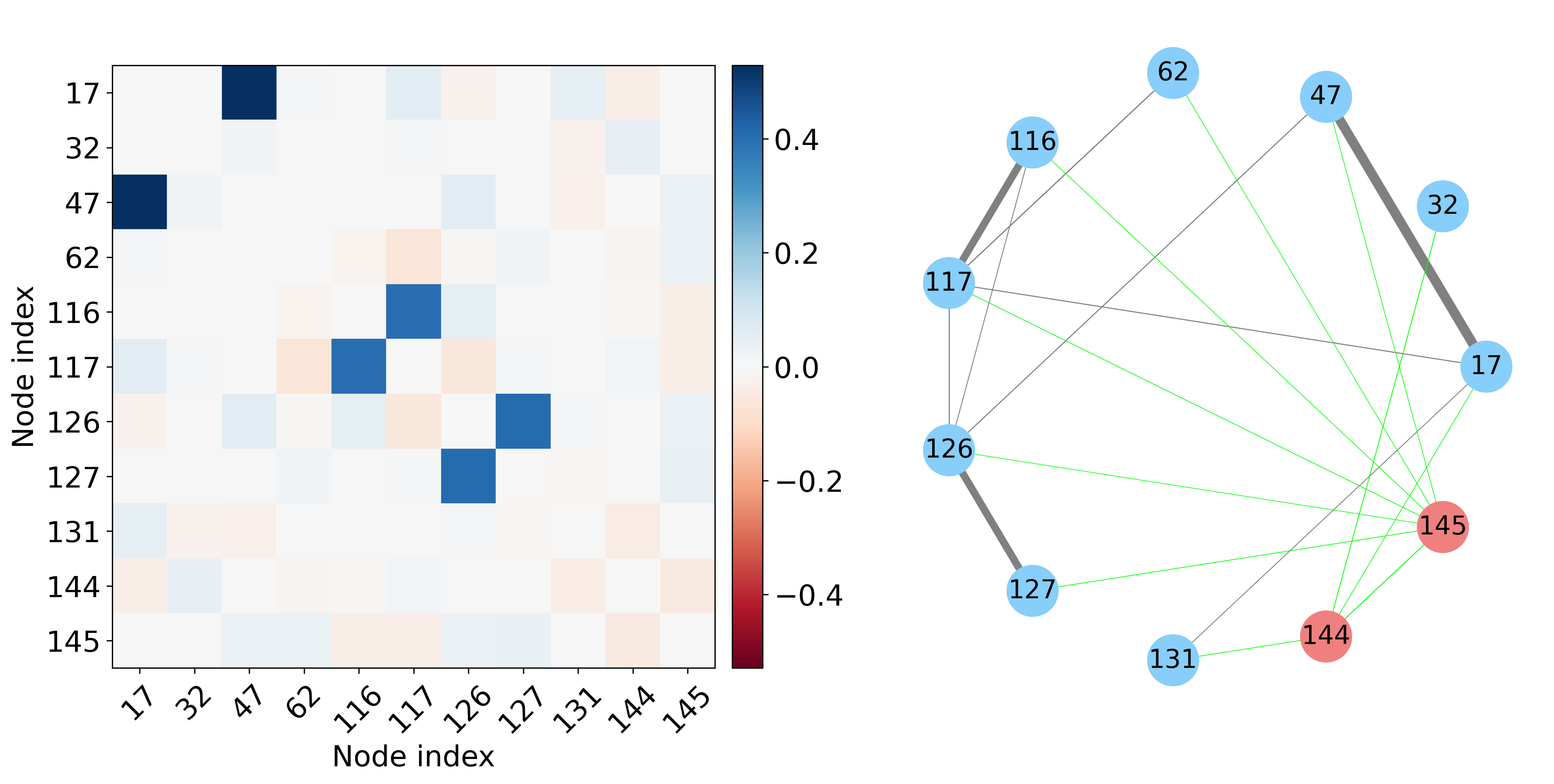}}
	\end{subfigure}
    \begin{subfigure}[Scenario 1: $\lambda=2.82$, with links between the outputs.]{\includegraphics[width=\columnwidth]{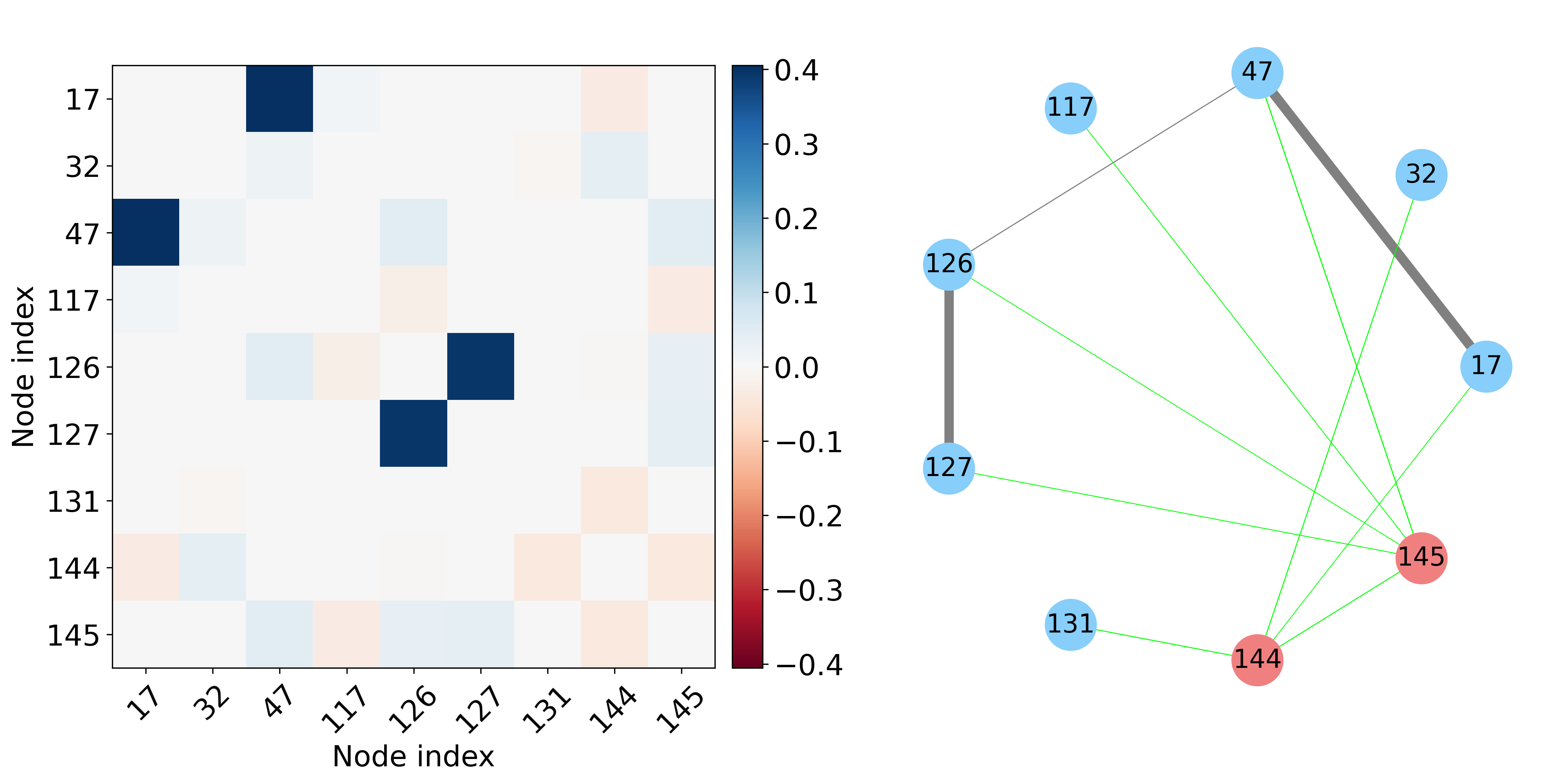}}
	\end{subfigure}
    \vspace{-3mm}
	\caption{{Graph representation and associated adjacency matrix for the ARAUS dataset described in Section~\ref{s:araus}. We consider Scenario~1, where the connection between the output variables (valence and arousal), corresponding here to nodes 144 and 145, is allowed. Panels (a) and (b) show the inferred structures for two values of $\lambda$.}}
    \label{fig:graphs_araus}
\end{figure*}


{We now apply the proposed method to the ARAUS dataset described in Section~\ref{s:araus} in order to analyze its graph structure. Similar to the EMO experiments, we focus on the connectivity between the two emotional outputs. ARAUS contains 143 acoustic features and 20022 observations, yielding the data matrix $\bbX \in \reals^{145 \times 20022}$, where the last two rows correspond to the output variables. We consider the setting described in Scenario~1, i.e., allowing links between the outputs and computing the error over all nodes.

The results obtained by Algorithm~\ref{alg:sem_guaed} are shown in Figure~\ref{fig:elbow_araus}. The elbow is located at 327 links, corresponding to $\lambda=2.82$, whereas the uncertainty interval ranges from 265 to 574 links, corresponding to $\lambda \in [10.81,0.56]$. Compared with the corresponding EMO result in Figure~\ref{fig:Picasso_comeme_el_casso3_2}(a), the interval obtained for ARAUS is wider. In particular, only two points lie outside the right boundary, suggesting that the error is less sensitive to the choice of $\lambda$. In addition, note that the elbow is close to the lower end of the interval, which implies that the selected value of $\lambda$ is close to the sparsest admissible graph. 
The corresponding graph representations are shown in Figure~\ref{fig:graphs_araus}, both for the elbow value ($\lambda=2.82$) and for the higher end of the interval ($\lambda=0.56$). 
Compared with the EMO analysis, the ARAUS graphs exhibit a larger number of weaker connections, suggesting that the influence of individual features on the outputs is more distributed. Consistent with the previous analysis of EMO dataset, we notice that the two output variables (nodes 144 and 145) are connected.
We also observe that the most relevant variables linked to the outputs are mainly frequency-domain features. Interestingly, node 144 is connected only to three other features (17, 32, and 131) in both cases, corresponding to psychoacoustic and frequency-domain variables. In contrast, node 145 is mainly related to time-domain and frequency-domain variables, being connected to nodes 47, 117, 126, and 127 at the elbow value, and to nodes 16, 62 for the smaller value of $\lambda$.}


\subsection{Previous results in the literature}

\begin{table*}[h!]
\begin{center}
\caption{Summary of current methods and models employed in the literature with EMO, along with the number of variables and the main ones. A: Arousal, V: Valence; M: Mean, Std: Standard Deviation.} \label{TableResultsLiterature1}
{
\begin{tabular}{|c|c|c|c|c|c|} 
 \hline
\multirow{2}{*}{Work}  & \multirow{2}{*}{Name of method}   & 
 \multicolumn{2}{c|}{ Number of variables}
 & 
  \multicolumn{2}{|c|}{ Main variable/s}\\ 
  \cline{3-6}
  &  & Arousal & Valence&  Arousal & Valence\\ 
 \hline
  \hline
 \cite{fan2017emo} & Support Vector Regression  &  39  & 39 & ---& ---  \\ 
 \cite{fan2018soundscape}   & Several Deep Learning methods & 54 & 54  & --- & --- \\ 
  \cline{5-6}
\cite{abri2020predicting}
  & Random Forest and other methods &  26   & 29  &
   \multicolumn{2}{|c|}{1, 2, 4}\\ 
   \cline{5-6}
 \cite{abri2021comparative}  & Random Forest &  15   & 14  & Roughness M, 50 & Roughness M, 1 \\
    \cline{5-6}
\cite{ntalampiras2020emotional}  & Convolutional Neuronal Networks &  23  & 23  &      \multicolumn{2}{|c|}{$\sim$ 24-36, 37-49}\\
  \cline{5-6}
 \cite{OurPaperSound}   &  Multiple Linear Regression &  7  &  16  & 113, 114 & 114, 113\\ 
   \cline{5-6}
  \cite{DTR_Access_24}     &   Decision Tree Regression &   2 &  4 & 113, 4 &  114, 102\\ 
  \cline{5-6}
\cite{serradilla2024emotional}     &   Convolutional Neuronal Networks &   64 &  64 & \multicolumn{2}{|c|}{audio spectrograms as images}\\
  \cline{5-6}
   {\bf Here}  &    {\bf SEM with LASSO} &    {\bf  7}&    {\bf 6}&     {\bf 113, 14}&    {\bf 114, 3 (and Arousal)}\\
 \hline
\end{tabular}
}
\end{center}
\end{table*}

The results of this study are consistent with those of \cite{OurPaperSound}, where a simple linear model was considered. With nonlinear models (such as decision trees \cite{DTR_Access_24}), a smaller number of features has been suggested compared to \cite{OurPaperSound}. The need for more variables for valence was also observed in \cite{abri2020predicting, abri2021comparative}. Table~\ref{TableResultsLiterature1} summarizes the results reported in other studies considering different regression methods. Moreover, Table~\ref{TableResultsLiterature1} presents the prediction model, the number of variables for the two outputs, and the {most} relevant variables reported in the literature. 
{In some previous works, the names of the selected variables were not explicitly specified.}
From Table~\ref{TableResultsLiterature1}, {we observe that psychoacoustic variables play a prominent role in previous studies based on the EMO dataset, such as MFCCs, Roughness, Fluctuation Strength, and Loudness.}
Predictive models based on recent soundscape databases also highlight the key role of psychoacoustic features. ARAUS showed that maximum loudness was the most important variable for ISO Pleasantness (i.e., the equivalent of valence in this work) \cite{ooi2023araus}, {followed by frequency-domain variables}. On the other hand, ISD proposed a model for ISO Pleasantness with Loudness (fifth percentile) as an extremely important variable, along with time-domain and frequency-domain variables (i.e., based on $L_{Aeq}$, $L_{Ceq}$, and sound pressure level percentiles). Moreover, ISO Eventfulness (i.e., the equivalent of arousal in this work) was modelled by considering other psychoacoustic variables (e.g., Fluctuation strength and Roughness). These ideas are in line with the conclusions of a recent review on soundscape modeling \cite{lionello2020systematic}. The soundscape models presented in this study result in a similar and parsimonious set of features {to those} used in the literature with other methods and datasets. 
{Our proposal also identifies psychoacoustic variables, particularly Loudness, as central, and complements them with frequency-domain and time-domain features.}

\section{Final discussion}\label{s:conclusions}

{In recent decades}, soundscape research has become one of the most active topics in acoustics. SER modelling benefits from feature selection by leading to more parsimonious and interpretable solutions. {Simpler models and identification of relevant features save time and resources in research problems.}
 
The proposed approach can be viewed as a combination of a wrapper method and a filter method for feature selection. Specifically, we employ a graph topology learning framework that allows the dependencies and connections among all variables, both inputs and outputs, to be analyzed jointly. {This enables to quantify the strength of the connections with the outputs (and thus build rankings, as in a wrapper method) and to observe and discard variables according to their correlations (as in a filter method).}
The topology of the graph is obtained by considering a {linear} SEM model (SEM) with LASSO penalization.

Furthermore, we have proposed an innovative procedure for handling the parameter $\lambda$
in LASSO. We introduced the G-UAED technique to obtain elbow points and intervals for $\lambda$. Thus, we can control the sparsity of the model. Benchmarking with previous studies using the EMO dataset confirmed a clear agreement on the most prominent features for SER modelling. {This work highlights psychoacoustic variables (e.g., 113 and 114) as the most relevant, while complementing them with time-domain and frequency-domain features (e.g., 14 and 3)}.
Nevertheless, our graph-based approach captures the structure between all features, providing new insights.
\begin{itemize}
    \item The two analyzed outputs, arousal and valence, appear to be very correlated, {which contrasts with some assumptions in classical SER theory}.
    \item {Valence requires fewer variables for modeling than arousal}. This finding is in contrast to those of previous studies.
    \item The suggested number of features required for arousal (i.e., 7) is in line with the results reported in recent studies in the literature.
    \item Relevant variables such as 20 and 14, or 4 and 3, 14 and 13, or 113 and 114, are highly correlated and, in many cases, can be considered interchangeable. Some results in the literature differ only with respect to these variables but are, in fact, in agreement.
\end{itemize}
Clearly, the proposed approach is applicable to other soundscape databases {and even to entirely different datasets or applications, as demonstrated with the ARAUS dataset}. 

\vspace{2mm}\noindent
{{\bf Future research.}
As part of future work, we plan to analyze the signs of the coefficients associated with the graph links, whereas the present ranking is based solely on their absolute values. Furthermore, directed graph structures can also be considered by relaxing the symmetry constraint on the matrix ${\bf A}$, i.e., removing the condition ${\bf A} = {\bf A}^{\top}$ and introducing a continuous acyclicity constraint. This formulation enables the estimation of directed acyclic graphs (DAGs) using continuous optimization techniques, which is a promising approach that warrants further investigation regarding algorithmic scalability. Such DAG structures may reveal potential causal relationships that, to the best of our knowledge, have not yet been explored in soundscape literature. A more ambitious long-term direction is to apply the proposed methodology to multiple datasets in order to develop a unified and parsimonious soundscape model capturing only the most relevant features.}

\bibliographystyle{IEEEtran}
\bibliography{myIEEEabrv,bibliografia}

\begin{IEEEbiography}[{\includegraphics[width=1in,height=1.25in,clip,keepaspectratio]{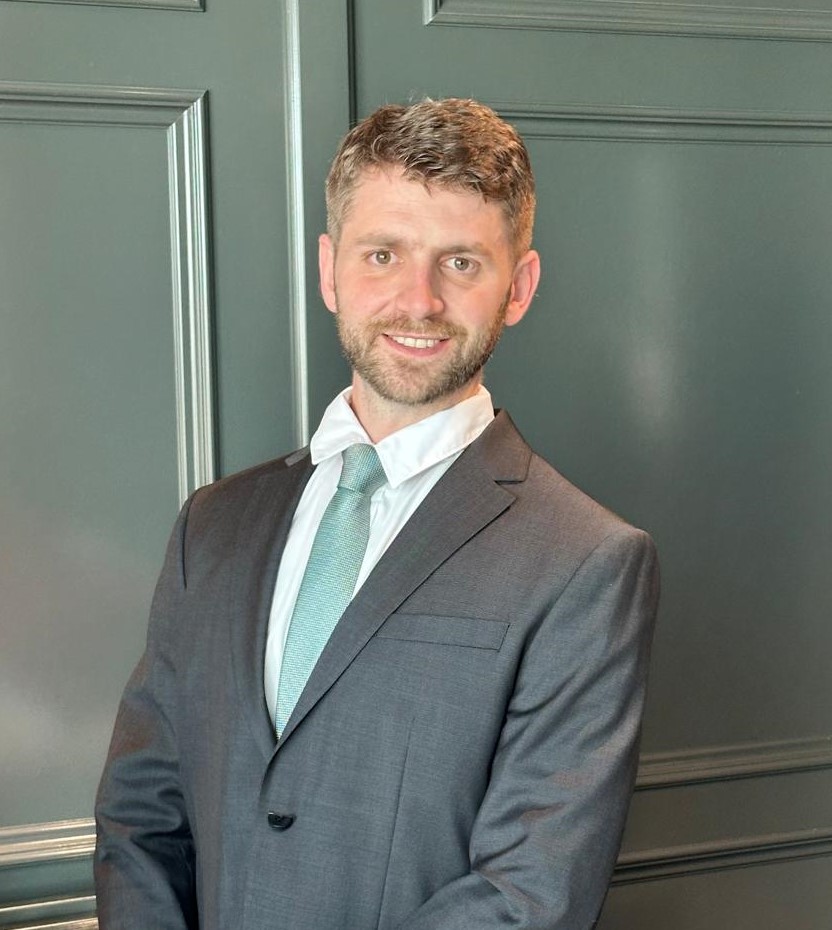}}]{Samuel Rey} (Member, IEEE)
received the B.Sc., M.Sc., and Ph.D. degrees in telecommunication engineering from King Juan Carlos University (URJC), Madrid, Spain, in 2016, 2018, and 2023, respectively, all with highest honors.
In 2023, he joined the Department of Signal Theory and Communications, URJC, where he is currently an Assistant Professor. 

His current research interests include graph signal processing, machine learning, nonconvex optimization, and data science over networks. Dr. Rey has served on the organizing committees of several international conferences. He received the Best Young Investigator Award from URJC in 2018 and was awarded the Spanish National FPU Scholarship for PhD studies in the same year.
\end{IEEEbiography}

\begin{IEEEbiography}[{\includegraphics[width=1in,height=1.25in,clip,keepaspectratio]{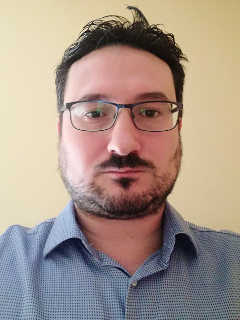}}]{Luca Martino} received the M.Sc. degree in electronic
engineering from the Politecnico di Milan,
Milan, Italy, in 2006, and the Ph.D. degree in statistical
signal processing from the Universidad Carlos
III de Madrid, Spain, in 2011.
He spent two years with the Department of
Statistics at the University of Helsinki, Finland.
He carried out  one year of Post-Doctoral Research at the S{\~a}o
Paulo Research Foundation (FAPESP), S{\~a}o Paulo,
Brazil, and  more than one year at the University of Valencia,
Spain. Since January 2020, he has been an Associate
Professor at the Universidad Rey Juan Carlos (URJC), Madrid, Spain.  Currently, He is  Associate Professor at Universit{\`a} di Catania, Italy.  He has published more than 70 journal articles from 2009 to 2023.
His main research interests include Monte
Carlo methods (importance sampling, MCMC, particle filtering), general problems and methods
for regression-filtering-smoothing, stochastic processes for regression (e.g., Gaussian processes), and multilabel classification.  He has been included in the ``Ranking of the World Scientists: World's Top 2\% Scientists'' since 2019, done by Researchers of Stanford University.
\end{IEEEbiography}

\begin{IEEEbiography}
[{\includegraphics[width=1 in,height=1.25 in,clip,keepaspectratio]{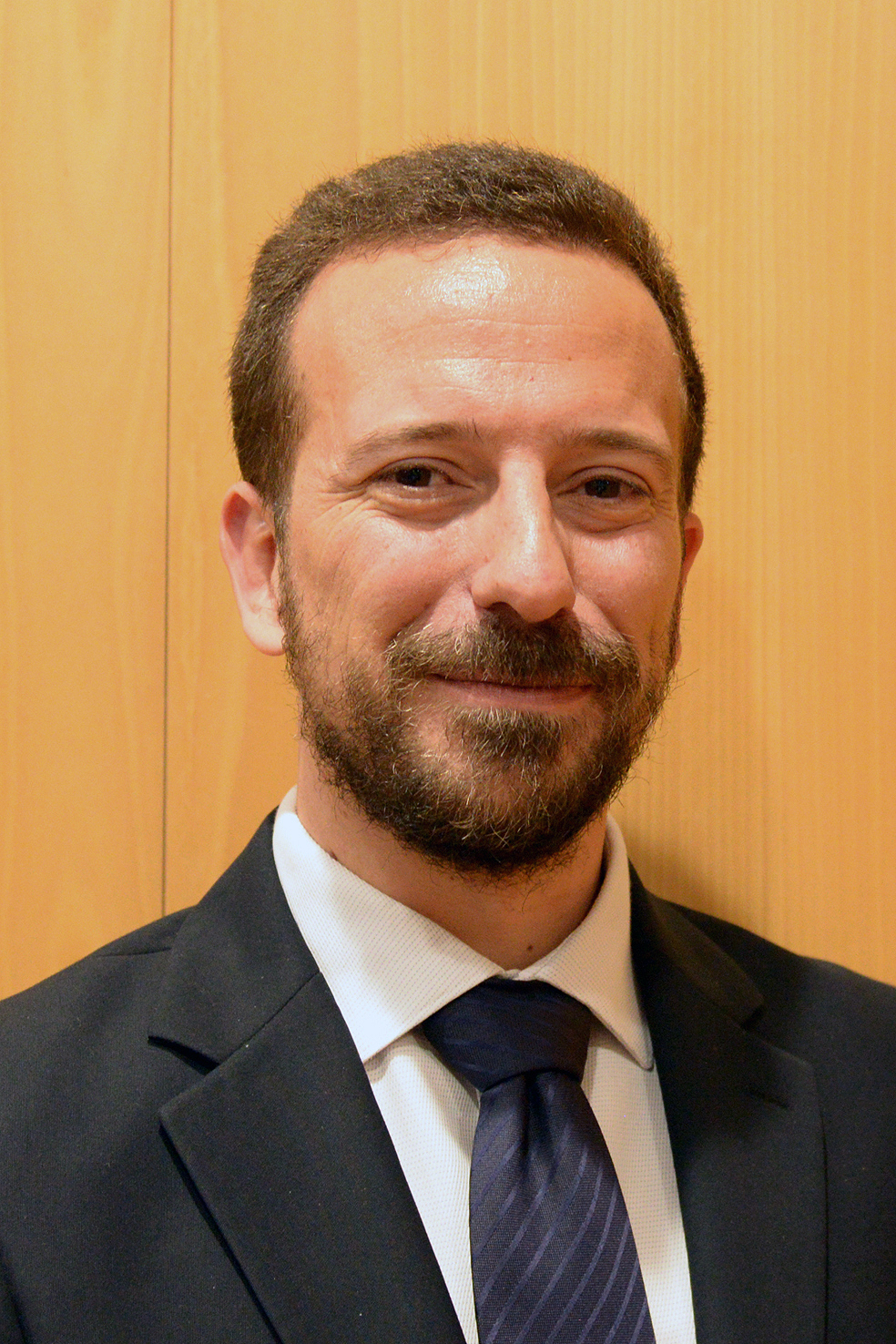}}]{\textbf{Eduardo Morgado}} received the M.Sc. degree in telecommunication engineering from the Carlos III University of Madrid, Leganés, Spain, in 2004, and the Ph.D. degree in telecommunication engineering from Rey Juan Carlos University, Fuenlabrada, Spain, in 2009, where he is currently an Associate Professor in the Department of Signal Theory and Communications. His research interests include signal processing for wireless communications with applications in ad hoc, sensor networks, biomedical engineering, and acoustics.
\end{IEEEbiography}

\begin{IEEEbiography}
[{\includegraphics[width=1in,height=1.25in,clip,keepaspectratio]{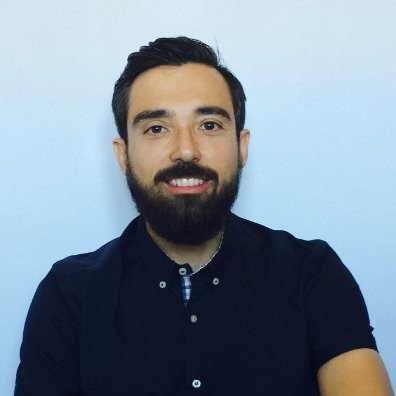}}]{\textbf{Roberto San Millán-Castillo}}
received
the B.Sc. degree in Telecommunications Engineering
(Electrical engineering), with a minor in
sound and image, from Universidad Politécnica
de Madrid (UPM), in 2000, the M.Sc. degree in
Project Management from Universidad Antonio
de Nebrija, in 2012, the M.Sc. degree in Acoustic
Engineering (research training) from UPM,
in 2013, and the Ph.D. degree in multimedia and
communications from UC3M+URJC in 2020.
In 2011, he joined Universidad Rey Juan Carlos (URJC), Spain, as a
Lecturer. Since 1999, he had hands-on experience in the industry as an
acoustics, noise control, audio, and instrumentation consultant in different
positions such as engineering, project management, and sales. All at once,
he taught an endless number of training courses at companies, universities,
Professional associations and government institutions. His main research
interests include signal processing and machine learning techniques applied
to practical problems with acoustical and audio signals, respectively.
\end{IEEEbiography}


\vfill

\end{document}